\documentclass{aa}
\usepackage{pifont}

\usepackage{graphicx}
\usepackage[varg]{txfonts}

\usepackage[hidelinks]{hyperref}

\newcommand{\zav}[1]{\left(#1\right)}
\newcommand{\hzav}[1]{\left[#1\right]}

\newcommand{\vecit}[1]{\ensuremath{\boldsymbol{#1}}}
\newcommand{\taucont}{\ensuremath{\tau_\text{cont}}}
\newcommand{\tauline}{\ensuremath{\tau_\text{line}}}
\newcommand{\taurand}{\ensuremath{\tau_\text{rand}}}
\newcommand{\cmf}{\ensuremath{\leadsto}}
\newcommand{\nucmf}{\ensuremath{\nu^\cmf}}
\newcommand{\rf}{\ensuremath{\sharp}}
\newcommand{\nurf}{\ensuremath{\nu^\rf}}
\newcommand{\vinfty}{\ensuremath{V_\infty}}
\newcommand{\Rinfty}{\ensuremath{R_\infty}}

\newcommand{\der}{\ensuremath{\mathrm d}}

\newcommand{\bolk}{\ensuremath{k}_\text{B}}

\newcommand{\Lstar}{\ensuremath{L_\mathrm{star}}}

\newcommand{\densnula}{\ensuremath{\varrho_{0}}}
\newcommand{\densclump}{\ensuremath{\varrho_\text{cl}}}

\newcommand{\nelec}{\ensuremath{n_\text{e}}}

\newcommand{\Telec}{\ensuremath{T_\text{e}}}

\newcommand{\Nlevel}{\ensuremath{\text{NL}}}
\newcommand{\totrate}{\ensuremath{\mathcal{R}}}
\renewcommand{\totrate}{\ensuremath{\mathcal{P}}}

\newcommand{\Lalpha}{\ensuremath{\text{Ly}\alpha}}
\newcommand{\Halpha}{\ensuremath{\text{H}\alpha}}

\newcommand{\anyl}{\ensuremath{f}}

\newcommand{\modgrid}{model grid}
\renewcommand{\modgrid}{modGrid}
\newcommand{\propgrid}{propagation grid}
\renewcommand{\propgrid}{propGrid}

\newcommand{\Nbasic}{\ensuremath{N^\text{B}}}

\newcommand{\jednad}{\mbox{1\protect\nobreakdash\,D}}
\newcommand{\dvad}{\mbox{2\protect\nobreakdash\,D}}
\newcommand{\trid}{\mbox{3\protect\nobreakdash\,D}}
\newcommand{\CMFGEN}{{\tt CMFGEN}}
\newcommand{\FASTWIND}{{\tt FASTWIND}}
\newcommand{\PoWR}{{\tt PoWR}}
\newcommand{\Tardis}{{\tt Tardis}}

\usepackage{tikz}
\usetikzlibrary{arrows,snakes,backgrounds}

\begin{document}

\title{Progress towards a 3D Monte Carlo radiative transfer code for outflow wind modelling}
\author{J. Fi\v{s}\'{a}k
        \inst{1,2}
        \and
        J. Kub\'{a}t
        \inst{2}
	\and
	B. Kub\'{a}tov\'{a}
	\inst{2}
	\and
	M. Kromer
        \inst{3}
	\and
	J. Krti\v{c}ka
	\inst{1}
	}
\institute{
Masaryk university, Faculty of Science,
Kotl\'{a}\v{r}sk\'{a} 2, Brno, Czech Republic
                \and
Astronomical Institute of the Czech Academy of Sciences, Fri\v{c}ova 298,
CZ-251 65, Ond\v{r}ejov, Czech Republic
\and
Heidelberger Institut f\"{u}r Theoretische Studien, Schloss-Wolfsbrunnenweg 35,
69118 Heidelberg, Germany
}

\abstract
{
Radiative transfer modelling of expanding stellar envelopes is an important
task in their analysis. To account for inhomogeneities and deviations from
spherical symmetry, it is necessary to develop a {\trid} approach to radiative
transfer modelling.
}
{
We present a {\trid} Monte Carlo code for radiative transfer modelling, which
is aimed to calculate the plasma ionisation and excitation state with the
statistical equilibrium equations, moreover, to implement photon-matter
coupling. As a first step, we present our Monte Carlo radiation transfer
routines developed and tested from scratch.
}
{
The background model atmosphere (the temperature, density, and velocity
structure) can use an arbitrary grid referred to as the {\modgrid}. The
radiative transfer was solved using the Monte Carlo method in a Cartesian grid,
referred to as the \propgrid. This Cartesian grid was created based on the
structure of the {\modgrid}; correspondence between these two grids was set at
the beginning of the calculations and then kept fixed. The {\propgrid} can be
either regular or adaptive; two modes of adaptive grids were tested. The
accuracy and calculation speed for different {\propgrid}s was analysed. Photon
interaction with matter was handled using the Lucy's macroatom approach. Test
calculations using our code were compared with the results obtained by a
different Monte Carlo radiative transfer code.
}
{
Our method and the related code for the {\trid} radiative transfer using the
Monte Carlo and macroatom methods offer an accurate and reliable solution for
the radiative transfer problem, and are especially promising for the inclusion
and treatment of {\trid} inhomogeneities.
}
{}
\maketitle

\section{Introduction}
The stellar wind is a kind of an outflow from the star that is stable on a
long-term scale. The stellar radiation field is often strong enough to
accelerate matter and create a stellar wind. Intensive radiation-driven stellar
winds are observed, for example, around O-type stars and Wolf-Rayet stars.
Stellar wind affects many astrophysical processes, including stellar evolution,
the chemical composition of galaxies, and the dynamics of interstellar matter.
A star can lose a significant amount of matter per unit of time; this amount is
characterized with a physical quantity named the `mass-loss rate', which is
important for the stellar evolution and for the galaxy chemical composition.
The accelerating mechanism of atoms by line radiation was suggested by
\cite{Milne1926}; line-driven winds were later described in detail by
\cite{Lucy1970}. The first hydrodynamical solution of the line-driven wind was
provided by \citet[also known as the CAK model]{Castor1975}. This model assumes
stationary, homogeneous, and spherically symmetric wind with a uniform outflow.

There is a wide range of existing codes for the stellar wind modelling. Stellar
winds of hot massive stars are routinely modelled assuming spherical symmetry
(i.e. \jednad), and taking into account the possibility of departures from the
local thermodynamic equilibrium (LTE). The latter approach is usually referred
to as NLTE. The codes solve the radiative transfer equation (hereafter RTE) and
the kinetic equilibrium equations (the NLTE line formation problem), typically
supplemented by the equation for temperature (either radiative equilibrium or
thermal balance). There exist three such widely used codes, namely {\CMFGEN}
\citep[e.g.][and references therein]{Hillier1987, Hillier1990,
Hillier_Miller_1998, Hillier_2012}, {\FASTWIND}
\citep[e.g.][]{Santolaya-Rey+1997, Puls+2005, Puls+2020}, and {\PoWR}
\citep[e.g.][]{Hamann2003, Hamann2004, Sander+2017}. In addition to these codes
used for analysis of observed spectra, other codes construct spherically
symmetric {\jednad} wind models using the solution of hydrodynamic equations
\citep{Pauldrach+2001, Krticka2017, sander2017, sundqvist2019}.

The assumptions of spherical symmetry and time independence were later shown
not to be correct. The spherical symmetry can be broken by the rotation of the
star \citep{Puls1993, Owocki1994, Petrenz2000}, by the magnetic field
\citep{UdDoula2014}, or by the accretion onto the compact object
\citep{Blondin1990, Feldmeiers1999}.

These processes are not the only ones which cause the wind asymmetry. Stability
analysis of stellar winds done by \cite{Carlberg1980}, \cite{Owocki1984},
\cite{Owocki2002}, \cite{Feldmeier1997}, and many others showed that density
and velocity are not monotonically changing functions. The corresponding
instability was found also by radiative hydrodynamical simulations in {\jednad}
\citep{Feldmeier1995,Owocki1988,Runacres2002} or in {\dvad}
\citep[e.g.][]{Dessart2003,Dessart2005,Sundqvist2018, Driessen2021}. The
inhomogeneities are called clumps. They can be introduced even into simplified
{\jednad} models using fiducial factors such as the clumping factor, or a
volume filling factor \citep[e.g.][]{Abbott1981}, or an effective opacity
formalism \citep{Sundqvist2018b}. However, there are possibilities to solve the
radiative transfer in clumped media precisely.

Consequently, there is a growing need for full {\trid} NLTE modelling of
massive star winds. Although there is a number of {\trid}
radiation-hydrodynamics codes for LTE atmospheric modelling
\citep[e.g.][]{Nordlund_Stein_2009, Ludwig_Steffen_2016, Freytag_etal_2019},
which were successful especially in the modelling of convection in cool stars,
related NLTE calculations are usually limited to the solution of the NLTE line
formation problem \citep[see][for a review]{Asplund_Lind_2010}. Nevertheless,
first steps towards full NLTE wind models have already been taken; however, due
to the complexity of the problem caused mainly by the necessity to include a
general velocity field, only simplified problems were solved. The key problem
is the way how the radiative transfer problem is treated. Several different
numerical methods have been applied to solve this problem. \cite{Adam_1990},
\cite{Lobel_Blomme_2008}, and \cite{Hennicker+2018} used the finite volume
method, while \cite{Papkalla_1995}, \cite{Korcakova_Kubat_2005},
\cite{Georgiev+2006}, \cite{Zsargo+2006}, \cite{Leenaarts_Carlsson_2009},
\cite{Ibgui+2013}, \cite{Stepan_TrujilloBueno_2013}, and \cite{Hennicker+2020}
used the short characteristics method. All these codes require enormous
computing time to solve even the simplest line formation problems.

An alternative to the solution of the multi-dimensional radiative transfer
equation by means of these methods which determine the intensity of radiation
is to use Monte Carlo (hereafter MC) radiative transfer \citep[see][for a
review]{noebauerSim2019}. This method is very useful for the optically thin
media such as circumstellar environments, and planetary nebulae. The main
advantage of the MC method is that the solution is calculated naturally in a
{\trid} space. This allows us to fully include some {\trid} phenomena such as wind
inhomogeneities, that is clumping.

There were some intermediate steps towards the {\trid} models.
\citet{Oskinova2004, Oskinova2006} and \cite{sundqvist2010} calculated
pseudo-{\dvad} models, later on followed by a pseudo-{\trid} model by
\cite{sundqvist2011}. There are several codes for the MC {\trid} RTE
calculation: HDUST (\citealt{Carciofi2004, Carciofi2006, Carciofi2017}), a
{\trid} NLTE radiative transfer code for Be stars, hot star winds, etc. The
PYTHON code (\citealt{Long2002}, extended by \citealt{Higginbottom2013}) is
capable of including analytical wind models, as well as disc, and supernova
models. It uses cylindrical or a spherical grid and it also can read in
arbitrary geometries. \citet{Surlan2012, Surlan2013} calculated wind clumping
with the MC code. They solved radiative transfer in doublet resonance lines.

Also the backward ray tracing method is being used -- this method can calculate
spectra based on the observer's position. This is useful for the calculation of
emergent radiation from non-symmetric objects. The standard Monte Carlo method
deals with photon propagation through a medium. Usually we observe a star via
detectors occupying a very small area. A very large amount of photons must be
propagated to create a spectrum without a significant noise. The ray tracing
method evades this disadvantage and solves RTE along specific emergent rays and
then we get spectra for the given direction.

There are also several MC codes calculating the radiative transfer through the
supernova ejecta. These codes calculate time-dependent spectra to take the
rapid variations of circumstellar ejecta into account. As an example, the code
{\Tardis} \citep{Kerzendorf2014, Vogl2019, kerzendorf_wolfgang_2022} enables
the fast calculation of supernova spectra. More sophisticated codes
\citep{Lucy2005, Kromer2009, Kromer2009b} also simulated radioactive reactions
in the ejecta, which provide another source of radiation. There are codes
calculating the radiation of the star with both the supernova ejecta and the
circumstellar disc.

We present the {\trid} radiative transfer MC code which we developed from the
scratch; the  code is suitable for outflow wind modelling. In
Sect.~\ref{Sec:MCRT} we describe the NLTE MC radiative transfer method which is
implemented in the code. The general code structure is present in
Sec.~\ref{sec:codeStruct}, while in Sec.~\ref{sec:grid_test}, we present the
results of testing the propagation grid. Physical processes in the outflow,
which are implemented in the code, are described in Sec.~\ref{Sec:physProc}.
Computational details are presented in Sec.~\ref{kapitola_implem}, while
comparison of our code with the {\Tardis} code is presented in
Sec.~\ref{Sec:testCase}. In Sec.~\ref{Sec:concl} we summarize and outline
future work.

\section{NLTE Monte Carlo radiative transfer}\label{Sec:MCRT}
Our radiative transfer code is based on Monte Carlo method of
\cite{Lucy2002, Lucy2003}. This method traces propagation of indivisible energy
quanta (energy packets) through matter. The energy packets can have form of
radiation energy packets (r-packets), internal energy packets (i-packets), and
kinetic energy packets (k-packets). The packets may change their form in
radiation-matter interactions.

The matter is quantized into macro-atoms. Detailed properties of energy packets
and macro-atoms are described in \cite{Lucy2002}.

\subsection{Radiative packets}
\begin{figure}
 \centering
 \includegraphics[width=.5\textwidth]{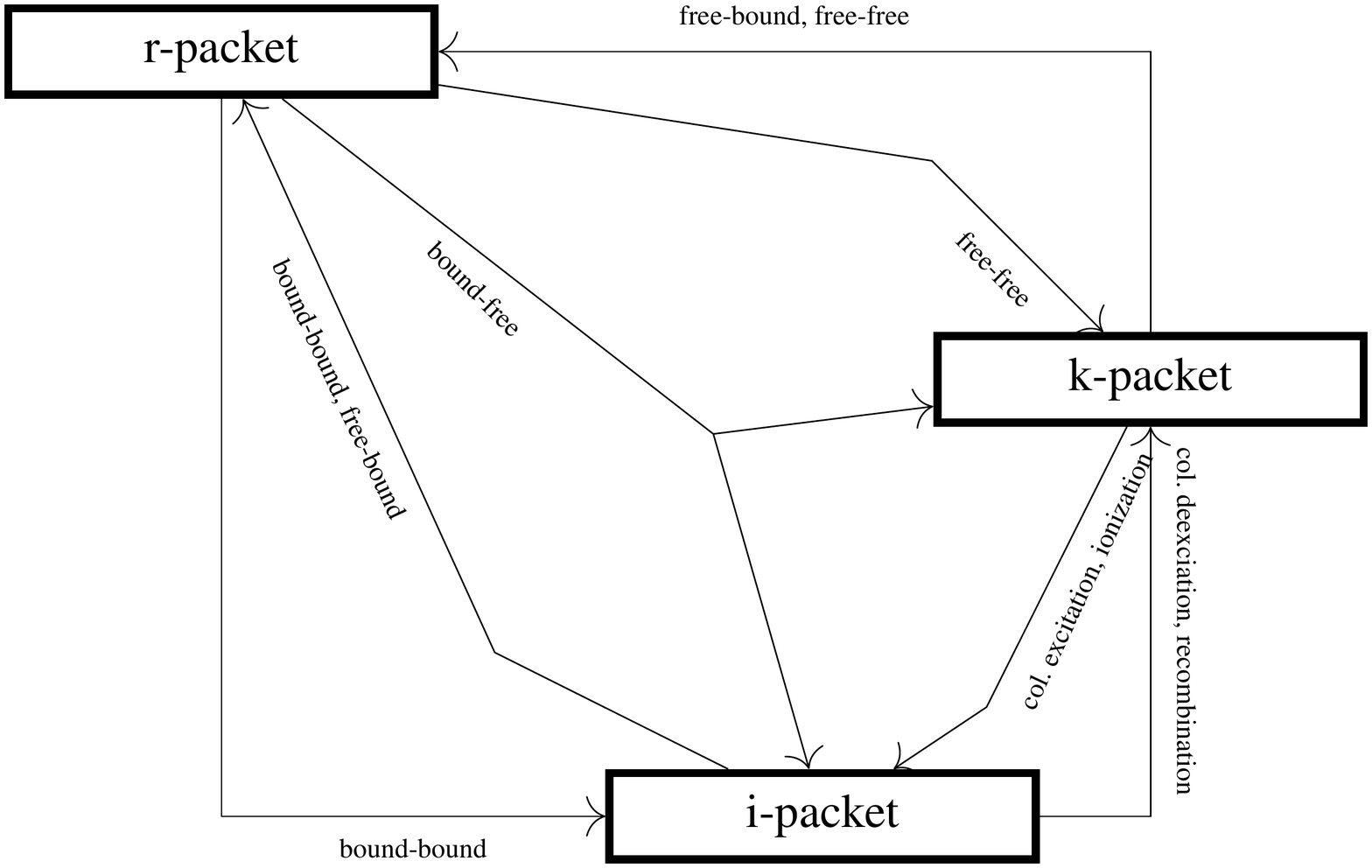}
 \caption{Scheme of the macro-atom interactions.}
 \label{Fig:procesy}
\end{figure}

Radiation in this method is quantized to radiation energy packets (r-packets),
which behave such as basic photonic quanta, but their energy is quite different
and does not correspond to a single photon, it is an ensemble of photons. The
propagation and changes of properties of these packets represent the radiation
energy transport through the atmosphere and the interaction of radiation with
matter. The radiation packets can be changed during radiation-matter
interaction (see Fig.~\ref{Fig:procesy}) to packets of internal energy
(i-packets) or kinetic (thermal) energy (k-packets).

Although the method is general and can be applied to many astrophysical
radiation sources, here we restrict ourselves to stars and circumstellar
regions. Since the Sobolev approximation is often used in supernovae modelling,
consequently, we adopted this approximation to enable easy comparison with
results obtained by supernova codes. Relaxation of this approximation in our
code will be discussed in a forthcoming paper.
\subsubsection{Creation of radiative packets}
\label{Sec:Creat-r-packets}
Creation of r-packets follows the lower boundary condition at the stellar
surface. Packet position, frequency, and direction have to be randomly chosen.
Starting positions of packets are chosen to be equally distributed across the
stellar surface. If we denote the stellar radius as $R_*$, the initial packet
position in the spherical coordinates is
\begin{subequations}
 \begin{equation}
  \boldsymbol{r}_\text{init} = (R_*\sin(\Theta)\cdot \cos(\Phi),
  R_*\sin(\Theta)\cdot \sin(\Phi), R_*\cos(\Theta)),
  \label{Eq:surfPos}
 \end{equation}
with
 \begin{equation}
  \Phi = 2\pi\tau,
  \label{Eq:spPhi}
 \end{equation}
 and
 \begin{equation}
  \cos(\Theta) = 2\sigma - 1.
  \label{Eq:spTheta}
 \end{equation}
Here $\tau$ and $\sigma$ are randomly generated numbers in the interval $[0,1]$
\citep[see also][Eqs. (26) and (27)]{noebauerSim2019}.
 \end{subequations}

The initial propagation direction of the r-packet is randomly chosen in the
outward direction from the stellar surface: another random unit vector
\begin{subequations}
 \begin{equation}
  \boldsymbol{n}_\text{init.} = (\sin(t)\cdot \cos(p), \sin(t)\cdot \sin(p),\cos(t)),
 \end{equation}
is generated with
 \begin{equation}
  p = 2\pi\tau',
 \end{equation}
 and
 \begin{equation}
  \cos^2(t) = \sigma'.
 \end{equation}
\end{subequations}
Here $\tau'$ and $\sigma'$ are randomly generated numbers in the interval
$[0,1]$ \citep[see also][Eq. (47)]{noebauerSim2019}. The propagation direction
in the stellar coordinate system is calculated as a double rotation (two
rotation-matrix multiplication), then the initial packet direction
$\boldsymbol{d}_\text{pack}$ is in the form
\begin{multline}
 \boldsymbol{d}_\text{pack} = (r_3\cdot \sin(t)\cdot \cos(p) + r_1\cdot \cos(t)\cdot \cos(p) - r_2 \cdot
 \sin(p),\\
 r_3\cdot \sin(t)\cdot \sin(p) + r_1\cdot \cos(t)\cdot \sin(p) + r_2 \cdot \cos(p),\\
 r_3 \cdot \cos(t) - r_1\cdot \sin(t)),
\end{multline}
where $r_i$ represent the Cartesian coordinates of the position vector
(\ref{Eq:surfPos}). The packet initial frequency follows the radiation
distribution emerging from the stellar surface, which may be the Planck
distribution or, more precisely, the calculated synthetic spectrum of the
stellar photosphere. Packets are created at the beginning of every iteration.
As the packet is not a photon, its energy must be established to ensure the
flux energy conservation. The packet energy $E$ (emitted per unit time) is
equal to
\begin{equation}
 E = \frac{\Lstar}{N},
\end{equation}
and is the same for all energy packets regardless their frequency. Here $N$ is
the total number of generated packets  and $\Lstar$ is the stellar luminosity.
\subsubsection{Propagation and interaction of radiative packets}
\label{Sec:Prop-r-packets}
The radiation energy packets (r-packets) are propagated through the wind. We
denote physical quantities expressed in the rest frame (the frame connected
with the centre of the star, hereafter RF) using the index $\rf$, and physical
quantities expressed in the co-moving frame (hereafter CMF) using the index
$\cmf$. The interaction of matter and radiation can occur only at a point where
the co-moving frame photon frequency $\nucmf$ equals the line transition
frequency ($\nucmf = \nu_\text{line}$).

Consider an r-packet with a rest frame frequency $\nurf$ at a position
$\boldsymbol{r}$ with density $\rho(\vecit{r})$ and temperature $T(\vecit{r})$.
We express the CMF frequency using the approximate Doppler shift formula
\begin{equation} \nucmf = \nurf\left(1 -
\frac{\boldsymbol{n\cdot \boldsymbol{v}}}{c}\right), \end{equation}
where $\boldsymbol{v}$ is the velocity vector of matter (in the RF) at a
given point $\vecit{r}$, and $\vecit{n}$ is the direction unit vector of a
light ray. As the packet propagates through the medium with a velocity
gradient, its CMF frequency is changing.

The continuum optical depth is calculated using the equation
\begin{equation}
 \label{eq:tauco}
 \taucont =
 \int\limits_{\mathbf{r}}^{\mathbf{r}_\text{R}}\,\text{d}\mathbf{r'}\cdot
 \chi_\text{cont}(\vecit{r'}),
\end{equation}
where $\boldsymbol{r}$ is the packet position, $\boldsymbol{r}_\text{R}$ is the
position of the closest possible line interaction (resonance point) of the
r-packet with matter and $\chi_\text{cont }(\vecit{r'})$ is the continuum
opacity,
\begin{equation}
\chi_{\text{cont}}(\vecit{r},\nucmf) =
\chi^{\text{Th}}(\vecit{r}) + \sum_{i=1}^{N_\text{I}}\chi^\text{ff}_i(\vecit{r},\nucmf)+
\sum_{i=1}^{N_\text{L}}\chi^\text{bf}_i(\vecit{r},\nucmf).
\end{equation}
The first term corresponds to Thomson scattering, the second one to free-free
processes,
and the third one to bound-free processes. $N_\text{I}$ is a total
number of all atomic ions and $N_\text{L}$ is a total number of energy levels
of all ions.

The line optical depth for a transition between a lower level $l$ and an upper
level $u$ integrated along a path $s$ from $0$ to $s_0$ is given by
\citep[see][Eq.\,4.15]{Kromer2009}
\begin{equation}
 \tauline = n_{l} \frac{B_{lu}h\nu}{4\pi}
  \left(1 - \frac{n_u g_l}{n_lg_u}\right)
  \int\limits_{\nucmf(s=0)}^{\nucmf(s=s_0)}\,
  \text{d}\nucmf\,\phi\left(\nucmf\right)
  \frac{\text{d}s}{\text{d}\nucmf},
  \label{eq:taulu}
\end{equation}
where $\phi$ is the line profile, $n_l, n_u$ are number densities of atoms at
levels $l$ (lower) and $u$ (upper), respectively, $g_l, g_u$ are corresponding
statistical weights, and $B_{lu}$ is the Einstein coefficient for a transition
from a level $l$ to a level $u$. In the Sobolev approximation this equation
simplifies to
\begin{multline} 
 \tauline = n_{l} \frac{B_{lu}h\nu}{4\pi}
  \left(1 - \frac{n_u g_l}{n_lg_u}\right)\\
  \times\left(\frac{\text{d}s}{\text{d}\nucmf}\right)_{\nu_{lu}}
  \times 
  \begin{cases}
   0 & \nu_{lu}\notin [\nucmf(0), \nucmf(s_0)],\\
   -1 & \nu_{lu}\in [\nucmf(0), \nucmf(s_0)],
  \end{cases}
\end{multline}
where $\nu_{lu}$ is the transition frequency. We note that the derivative
$(\mathrm{d}s/\mathrm{d}\nucmf)_{\nu_{lu}}$ is negative.
If we assume spherically symmetric velocity fields, this expression simplifies
(\citealt{Castor1974}, see also \citealt[Eq.\,22]{noebauer2015}) to
\begin{equation}
 \tauline = \frac{\kappa \rho c}{\nu_{lu}}
 \left[\mu^2\frac{\text{d}v}{\text{d}r} + \zav{1 - \mu^2} \frac{v}{r}\right]^{-1},
 \label{eq:tauluss}
\end{equation}
where $\mu$ is angle cosine between $\vecit{n}$ and $\vecit{v}(\vecit{r})$, and
\begin{equation}
\kappa \rho = \frac{\pi e^2}{m_\text{e}c}f_{lu} n_l
 \left(1-\frac{n_u}{n_l}\frac{g_l}{g_u}\right),
\end{equation}
according to \citet[Eq. (7)]{noebauer2015}. Here $f_{lu}$ is the oscillator
strength for a transition from a level $l$ to a level $u$.


In Monte Carlo radiative transfer, each r-packet is allowed to travel a
random optical depth $\taurand$ and then it interacts. This optical depth is
\citep[see, e.g.][]{whitney2011}
\begin{equation}
\label{eq:taura}
\taurand = - \ln(1 - z) = - \ln(z'), \quad z, z'\in [0,1],
\end{equation}
where $z$ is a random number. We can decide which type of the process (line or
continuum) happens by comparison of three optical depths, random optical depth
{\taurand} (Eq.~\ref{eq:taura}), line optical depth {\tauline}
(Eq.~\ref{eq:tauluss}), and continuum optical depth {\taucont}
(Eq.~\ref{eq:tauco}). The line and continuum optical depths are calculated by
integration along the photon path. The continuum optical depth increases
continuously, while the line optical depth in the Sobolev approximation changes
by jumps at particular resonance points. If
\begin{equation}
 \label{eq:letidal}
 \taurand \ge \taucont + \tauline,
\end{equation}
the photon continues its path, otherwise it interacts. The jump in line optical
depth caused by the Sobolev approximation enables a simplification of
determination whether a line or continuum transition happens. The type of
interaction (line or continuum) is chosen according to a simple rule \citep[as
in][]{Kromer2009}, it corresponds to the first transition which caused
invalidity of Eq. \eqref{eq:letidal}.

If a line transition happens, then the specific line is chosen according to the
scheme mentioned above, namely the one which caused invalidity of
\eqref{eq:letidal} is chosen.

If a continuum process is chosen, it has to be decided (using random numbers)
which specific process occurs. Currently, bound-free and free-free transitions,
and scattering on free electrons (Thomson scattering) are taken into account.
The Thomson scattering causes reemission of a new packet in a random direction
(here we assume isotropic scattering on free electrons) with the same CMF
frequency as of the original packet. The CMF quantities are recalculated to
corresponding RF values. A free-free process can transform an r-packet into
thermal kinetic energy (a k-packet). Bound-free processes contribute to both
kinetic and internal energy (k- and i-packets, respectively). Following
\citet[][Eq.~(27)]{Lucy2003}, the packet converts to an i-packet with a
probability $w_\text{i} = \nu_{i}/\nucmf$, where $\nu_{i}$ is the ionization
edge frequency. The packet is transformed to the k-packet otherwise
($w_\text{k} = 1 - \nu_{i}/\nucmf$).

Free-free processes transform an r-packet into a k-packet and vice versa.
Bound-free processes contribute to both internal energy and kinetic energy. For
each r-packet we decide (again using random numbers) whether it changes to a
k-packet or to an i-packet. The line event changes the r-packet to an i-packet
(macro-atom activation).

\subsection{Macro-atom statistical equilibrium equations}
\label{kap_makroatom}
Let us write the kinetic (statistical) equilibrium equation for the level $i$
\citep[see][Chapter 14]{Hubenyc2015}
\begin{equation}\label{Eq:esezaklad}
 \sum_{j\neq i}^\Nlevel n_i(R_{ij} + C_{ij}) =
  \sum_{j\neq i}^\Nlevel n_j(R_{ji} + C_{ji}),
\end{equation}
where $R_{ij}$ represents the radiative rates from the atomic state $i$ to the
atomic state $j$, $C_{ij}$ the collisional rates, and {\Nlevel} denotes the
total number of levels for the given atom (including all atomic ions). Let us
assume that the levels are ordered with increasing excitation energy, $\forall
i: \varepsilon_i < \varepsilon_{i+1}$ starting with the ground level of the
neutral atom and ending with the fully ionized atom (with only one energy
level). Energy of a level is equal to the sum of ionization energy of an ion
and excitation energy of the state. According to \cite{Lucy2002}, the sums in
Eq.~\eqref{Eq:esezaklad} can be split into two parts: transitions to levels
with excitation energies either lower or higher than the excitation energy of
the level $i$,
\begin{multline}
 \sum_{l= 1}^{i-1} n_i(R_{il} + C_{il}) +
 \sum_{u= i + 1}^\Nlevel n_i (R_{iu} + C_{iu})=\\ 
 \sum_{l = 1}^{i - 1} n_l (R_{li} + C_{li}) +
 \sum_{u = i + 1}^\Nlevel n_u (R_{ui} + C_{ui}).
\end{multline}
We define the total rate for the transition $i \rightarrow l$ as $\totrate_{il}
= n_i(R_{il} + C_{il})$, after a simple manipulation we get
\cite[see][Eq.~4]{Lucy2002}:
\begin{equation}
 \sum_{l= 1}^{i-1} \zav{\totrate_{li} - \totrate_{il}} +
 \sum_{u= i + 1}^\Nlevel \zav{\totrate_{ui} - \totrate_{iu}}
 = 0.
 \label{Eq:see}
 \end{equation}
It is possible to rewrite this equation in the form of energy absorbed and
emitted by processes connected with the level $i$
\citep[see][Eq.\,5]{Lucy2002}. To this end, we define the unit volume energy
rates: $\dot E_i^\text{R}$ (radiative energy emission rates) and $\dot
E_i^\text{C}$ (collisional energy emission rates), and the absorption energy
rates $\dot A_i^\text{R}$ (radiative) and $\dot A_i^\text{C}$ (collisional).
These terms express energy absorbed or emitted per unit time and unit volume.
The radiative energy rates can be expressed as
\begin{subequations}\label{AEdef}
\begin{equation}
 \dot A_i^\text{R} = \sum_{l=1}^{i-1}n_l R_{li}(\varepsilon_i -
 \varepsilon_l),\;\;
 \dot E_i^\text{R} = \sum_{l=1}^{i-1} n_i R_{il}(\varepsilon_i -
 \varepsilon_l),
 \label{Eq:er:aeraddown}
\end{equation}
the collisional energy rates as
\begin{equation}
 \dot A_i^\text{C} = \sum_{l=1}^{i-1}n_l C_{li}(\varepsilon_i - \varepsilon_l),\;\;
 \dot E_i^\text{C} = \sum_{l=1}^{i-1} n_i C_{il}(\varepsilon_i -
 \varepsilon_l).
 \label{Eq:er:aecoldown}
\end{equation}
\end{subequations}
Let us express the sum of these rates using total transition rates $\totrate$,
\begin{subequations}\label{AEdefsum}
\begin{equation}
  \dot A_i^\text{R} +  \dot A_i^\text{C} =
   \sum_{l=1}^{i-1} \totrate_{li}\varepsilon_i -
    \sum_{l=1}^{i-1}\totrate_{li}\varepsilon_l,
\end{equation}
\begin{equation}
  \dot E_i^\text{R} +  \dot E_i^\text{C} =
   \sum_{l=1}^{i-1} \totrate_{il}\varepsilon_i -
    \sum_{l=1}^{i-1}\totrate_{il}\varepsilon_l.
\end{equation}
\end{subequations}
Multiplying \eqref{Eq:see} with $\varepsilon_i$ and substituting for $\sum_{l= 1}^{i-1}
\zav{\totrate_{li} - \totrate_{il}}\varepsilon_i,$ from \eqref{AEdefsum} we get
\begin{multline}
 \dot E_i^\text{R} +\dot E_i^\text{C} +
\sum_{u= i + 1}^\Nlevel \totrate_{iu}\varepsilon_i +
\sum_{l= 1}^{i-1} \totrate_{il}\varepsilon_l =\\
\dot A_i^\text{R} +\dot A_i^\text{C} +
\sum_{l= 1}^{i-1} \totrate_{li}\varepsilon_l +
\sum_{u= i + 1}^\Nlevel \totrate_{ui}\varepsilon_i,
\label{Eq:esema}
\end{multline}
which is equivalent to Eq.~\eqref{Eq:see}. Terms with $\totrate_{ij}$ in
(\ref{Eq:esema}) describe energy flows between individual states. The equation
\eqref{Eq:esema} deals with macroscopic energy flow rates in a volume element.
Flows can be quantized into indivisible energy packets and atoms in the volume
element can be taken as a macro-atom with discrete energy.
\subsection{Internal energy packets}

Absorption of the radiation energy packet (the r-packet) via excitation or
ionization activates the macro-atom. Internal energy packets (i-packets)
represent excitation energy in excited atoms. Every i-packet must contain
information about atom(s) it represents: an atomic number, an ion number, and a
level index. The ion number and the level index may change during the i-packet
processing. To describe particular processes changing excitation and ionization
state, we use following index notation: $i$ -- the current level, $l$ -- a
lower level, $u$ -- an upper level, $m$ -- a level in a lower ionization state
than the level $i$, and $p$ an upper level ionization state. If necessary, we
specify the level with a lower index (E -- element, I -- ion, L -- level) and
its lower index classifies its concrete value, for example $i_{{I_k}{L_0}}$
represents the basic level of the ion k (the state is denoted as $i$).

Let us denote the total energy loss (left hand side of Eq.\,\ref{Eq:esema})
caused by transitions from the energy level $i$ as \citep[see][Eq.
4.38]{Kromer2009}
\begin{equation}
\dot{E}_i^\text{tot} = \dot{E}_i^\text{R} + \dot{E}_i^\text{C} + \dot{E}_i^\text{int, down} +
 \dot{E}_i^\text{int, up},
\end{equation}
where $\dot{E}_i^\text{R}$ and $\dot{E}_i^\text{C}$ denote radiative and
collisional deactivation (introduced in Eq.\,\ref{AEdef}), respectively, and
$\dot{E}_i^\text{int, up}$ and $\dot{E}_i^\text{int, down}$ mean internal
upward and downward jumps, respectively.

The macro-atom processing is probabilistic. Every possible process occurs with
a probability $p_{i\anyl}$, where $i$ denotes the initial (level, ion, element)
and $\anyl$ the final atomic state (again: level, ion, element).
Here $\anyl$ stands for $l$, $u$, $m$, or $p$, as introduced above. The internal
processes occur until the macro-atom is deactivated by one of the non-internal
(deactivation) processes, namely%
\footnote{Detailed expressions for radiative
($R$) and collisional ($C$) rates can be found in the Appendix
\ref{ap:radcolrates}.}
\begin{subequations}\label{i_deact}
 radiative deexcitation from the state $i$ to the state $l$ -- the probability of
       this process is equal to:
       \begin{equation}\label{rad_deexc}
        p^\text{R}_{il} = n_iR^\text{S}_{il}
        \left(\varepsilon_{i} - \varepsilon_{l}\right) / \dot{E}^\text{tot}_i,
       \end{equation}
  where $R^\text{S}_{il}$ is a rate in the Sobolev approximation, denoted with an upper index
  S,
  radiative recombination (considered only from an ion ground level):
  \begin{equation}\label{rad_reko}
   p^\text{R}_{im} = n_iR_{im}
   \left(\varepsilon_{i_{\text{L}_0}} - \varepsilon_{m} \right) / \dot{E}^\text{tot}_i,
  \end{equation}
  $R_{im}$ is the rate for recombination \eqref{Eq:ap:bfdown}, the index
  $i_{\text{L}_0}$ means the lowest level of the ion,
 collisional deexcitation -- in this case:
  \begin{equation}\label{col_deexc}
   p^\text{C}_{il} = n_iC_{il}
   \left(\varepsilon_{i} - \varepsilon_{l}\right) / \dot{E}^\text{tot}_i,
  \end{equation}
  $C_{il}$ is the collisional deexcitation rate 
  (see Section \ref{sec:colrat}), and
 collisional recombination (considered only from an ion ground level):
  \begin{equation}\label{col_reko}
   p^\text{C}_{im} = n_iC_{im}
   \left(\varepsilon_{i_{\text{L}_0}} - \varepsilon_{m}\right) / \dot{E}^\text{tot}_i,
  \end{equation}
  where $C_{im}$ is the collisional recombination rate
  (see Section \ref{sec:colrat}).
\end{subequations}

\begin{figure}
 \centering
 \includegraphics[width=200pt]{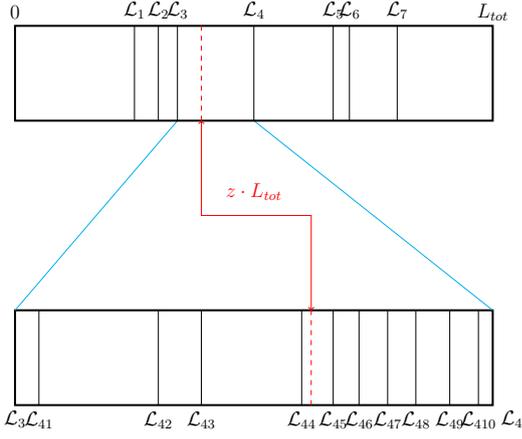}
 \caption{Scheme of choosing the corresponding process. There is saved
  information for concrete transitions (lower rectangle represents transition
  of a specific type with a definition (\ref{Eq:Lix})). Number of these
  transitions can be very high. For saving a computer time, first of all it is
  determined which process occurs. All total transition probabilities for
  specific processes are determined with Eq.~(\ref{Eq:Li}). After the process
  determination the programme goes through all possible transitions and chooses
  the right one which satisfies the condition described in the text. In the
  example here eight processes are possible, out of them the fourth process is
  chosen and it contains eleven transitions. The fifth transition is then
  selected.}
 \label{Fig:choice}
\end{figure}

The internal processes (which do not deactivate the macro-atom) taken into account include%
\begin{subequations}\label{i_intproc}
internal downward jump within the current ionization state: 
  \begin{equation}
   p^\text{int, down}_{il} =
   n_i\left(R^\text{S}_{il} + C_{il}\right)
   \varepsilon_{l} / \dot{E}^\text{tot}_i,
  \end{equation}
 internal downward jump to the lower ionization state:
  \begin{equation}
   \label{Eq:intdjm}
   p^\text{int, down}_{im} = 
   n_i\left(R_{im} + C_{im}\right)
   \varepsilon_{m} / \dot{E}^\text{tot}_i,
  \end{equation}
 internal upward jump within the current ionization state:
  \begin{equation}
   p^\text{int, up}_{iu} =
   n_i\left(R^\text{S}_{iu} + C_{iu}\right)
   \varepsilon_{i} / \dot{E}^\text{tot}_i,
   \label{Eq:inup}
  \end{equation}
  where $R^\text{S}_{iu}$ is the rate of transition to the upper level, and,
  finally, internal upward jump to a higher ionization state:
  \begin{equation}
   \label{Eq:inupp}
   p^\text{int, up}_{ip} =
   n_i\left(R_{ip} + C_{ip}\right)
   \varepsilon_{i} / \dot{E}^\text{tot}_i.
  \end{equation}
\end{subequations}
Equations \eqref{i_deact} and \eqref{i_intproc} contain rates which are not
corrected for stimulated emission. If we want to use the rates corrected for
the stimulated emission, we replace rates with Eqs.~\eqref{Eq:ab-stim:bbup} and
\eqref{Eq:emspont:bbdown}. By analogy, recombination can be treated as negative
photoionization via the Eq.~\eqref{Eq:photiontilde}. There are no downward
internal jumps to the ground level of the lowest ionization stage (the neutral
atom), since the energy $\varepsilon_l$ of this level is zero. \paragraph{The
choice of a particular process} The basic algorithm is shown in the
Fig.~\ref{Fig:choice}). Let us denote the total rate of all transitions from
the $i$-th level as
\begin{equation}
 L_i^\text{type} = \sum_\text{j}p_{ij}^\text{type},
 \label{Eq:Li}
\end{equation}
where $\text{type}$ has a value from the set rad deexc., rad. recomb., col
deexc, col recomb, internal: upward and downward jump within or to lower/upper
ionization state. Let us number the variable type as integers running from 1 to
the maximal number of processes included. Based on this definition of $L_i$ we
define the cumulative rate
\begin{equation}
 \mathcal{L}_k = \sum_\text{type=1}^k L_i^\text{type}
 \label{Eq:Lj}
\end{equation}
used in the upper panel of Fig.~\ref{Fig:choice}. Clearly $\mathcal{L}_0 = 0$.
We can also define a cumulative probability for a transition
\begin{equation}
 \mathcal{L}_{i,x} = \mathcal{L}_{i-1} + \sum_{y=1}^x p_{iy}^\text{type}.
 \label{Eq:Lix}
\end{equation}
Here $x$ is an index of the last transition taken into account. We generate a
random number $z \in [0, 1]$ (with the uniform distribution). The random number
$z$ is then multiplied with the total rate $L_\text{tot} = \mathcal{L}_N$
where $N$ is the total number of types of
processes ($N\le 8$, cf. Eqs.~\ref{i_deact} and \ref{i_intproc}). Then we check
which interval $(\mathcal{L}_{q-1}, \mathcal{L}_{q}]$ includes the number
$z\cdot L_\text{tot}$: the condition $\mathcal{L}_{q-1}\le z\cdot L_\text{tot}
< \mathcal{L}_q$ (where $q\in\{1, 2, \dots, N\}$ and $\mathcal{L}_0 = 0$) is
tested whether it is satisfied. For this selected index $q$ we search in a
similar way for an index $x$ satisfying the relation $\mathcal{L}_{q,x-1}\le
z\cdot L_\text{tot} < \mathcal{L}_{q, x}$ and these indexes $q$ and $x$ are
denoted as $Q$ and $X$, respectively.
This two-step process is faster than a one-step process which goes through
all transitions once and determines directly a specific transition.

When the radiative deexcitation in line \eqref{rad_deexc} or the radiative
recombination occurs, the i-packet transits into the r-packet. We have to
calculate its new frequency and direction. The direction unit vector is sampled
before the calculation of rates using the equation
\begin{equation}\label{eq:rpsmer}
 \boldsymbol{n} = (\sin(\Theta)\cdot \cos(\Phi), \sin(\Theta)\cdot
 \sin(\Phi),\cos(\Theta)),
\end{equation}
where $\Theta$ and $\Phi$ are determined using Eqs.~\eqref{Eq:spTheta} and
\eqref{Eq:spPhi}, respectively.

The rates depend on the Sobolev optical depth via  Eqs.~(\ref{Eq:absSob}) and
(\ref{Eq:emiSob}), see also \citet{KleinCastor1978}. For spherical
non-homological velocity fields $\mu$ does not vanish and
Eq.~(\ref{eq:tauluss}) depends on $\mu$.

If the radiative deexcitation (RD) occurs a new packet frequency must be
calculated in the following way: We find all possible downward transitions from
the state $i$. The probability of choosing $l$-th line is given by
\begin{equation}
 w_{il}^\text{RD} = \frac{p^\text{R}_{il}}{\sum_{l'} p^\text{R}_{il'}}.
\end{equation}
The next step is a choice of a process similar to the lower part of the
Fig.~\ref{Fig:choice}. The chosen line sets up the CMF frequency of a new
created r-packet. Currently a $\delta$\nobreakdash-function is assumed as a
line profile. The r-packet direction is sampled randomly according to
\eqref{eq:rpsmer}. The sampled frequency is connected to the CMF and it must be
transformed to the RF.

The probability of choosing a specific radiative recombination transition
$w_{im}^\text{RR}$ from those introduced by \eqref{rad_reko} is defined
similarly to the case of radiative deexcitation, namely
\begin{equation}
 w_{im}^\text{RR} = \frac{p^\text{R}_{im}}
  {\sum_{m'} p^\text{R}_{im'}}.
\end{equation}
The only difference between the treatment of radiative deexcitation and
radiative recombination is the sampling of a new packet frequency. In the case
of radiative recombination the frequency is sampled from the equation
\begin{equation}
 \int\limits_{\nu^\cmf}^\infty\,\text{d}\nu
\left(\eta_{im}^\text{fb}(\nu)\right)^\cmf =
 z\int\limits_{\nu_{i}}^\infty\,\text{d}\nu
 \left(\eta_{im}^\text{fb}(\nu)\right)^\cmf,
 \label{Eq:bffreq}
\end{equation}
where $z\in[0,1]$ is chosen randomly, the emissivity is given by
Eq.~(\ref{ap:bfemis}), and $\nu_i$ is the ionization frequency. The method of
sampling of a frequency is following: first the integral on the right hand side
is calculated. Second, the left hand side integral is repeatedly evaluated as a
function of its lower boundary until the integral value is approximatelly the
same as the right hand side. The lower integration boundary value {\nucmf} is a
new r-packet frequency. 

In the case of collisional deexcitation \eqref{col_deexc} and recombination
\eqref{col_reko}, the i-packet transits into a k-packet. No determination of
packet frequency and direction is needed in this case.

In the case of internal jumps only new actual state of macro-atom is set. New
transition probabilities \eqref{i_deact} and \eqref{i_intproc} are calculated
in the current new state and the process is repeated.

\subsection{Kinetic energy packets}
Kinetic energy packets represent free electrons in the medium. These packets do
not contain any information about their status and they are assumed not to move
in space. The only thing that has to be done is a choice of their interaction.
The total collisional cooling rate is equal to \citep{Kromer2009}
\begin{equation}
 \mathcal{C} = \sum_{i=1}^{\mathcal{N}_\text{I}}
  n_i\mathcal{C}_{i}^\text{ff}
  + \sum_{i=1}^{\mathcal{N}_\text{L}}
 \left(
  n_i\mathcal{C}_{im}^\text{fb,sp} +
  n_i\mathcal{C}_{ip}^\text{ion} +
  \sum_{i'=i+1}^{\mathcal{N}_\text{L}}
  n_i\mathcal{C}_{ii'}^\text{exc}
 \right),
\end{equation}
where $\mathcal{N}_\text{I}$ is the total number of ions (of element $k$), and
finally $\mathcal{N}_\text{L}$ is the total number of levels (of ion $j$ of
element $k$). Individual processes leading to a change of a k-packet include
 collisional excitation
  \begin{equation}\label{k_colexc}
   \mathcal{C}^\text{exc}_{iu} =
   n_i C_{iu}\left(\varepsilon_{u} - \varepsilon_{i}\right),
  \end{equation}
  where $C_{iu}$ is given by (\ref{eq:colbb}),
 free-free emission
  \begin{equation}\label{k_ff}
   \mathcal{C}^\text{ff}_{i} = C_0 q_{i}^2 T_\text{e}^{1/2}N_{i}n_\text{e},
  \end{equation}
  where $C_0 = 1.426\times 10^{-27}$ in CGS units, $q_{i}$ is the charge of ion
  $i$, and $N_i$ is the ion $i$ concentration,
free-bound transition (radiative recombination)
  \begin{equation}\label{k_fb}
   \mathcal{C}^\text{fb, sp}_{mi} =
   N_{m}n_\text{e}\left(\alpha^\text{E, spont}_{i} -
   \alpha^\text{spont}_{i}\right)
   \left(\varepsilon_{m_{\text{L}_0}} - \varepsilon_{i}\right),
  \end{equation}
  the cross section $\alpha^\text{E, spont}_{i}$ can be found in the
  Eq.~(\ref{ap:fbEspont}). The first term represents thermal and ionization
  energy converted to radiant energy,
  and the second term the ionization energy
  spontaneously converted to radiant energy in a given f-b transition.
 Finally, collisional ionization
  \begin{equation}\label{k_colion}
   \mathcal{C}_{im}^\text{ion} =
   c_{im}
   \left(\varepsilon_{m_{\text{L}_0}} - \varepsilon_{i}\right),
  \end{equation}
   where $c_{im}$ is written in the Eq.~(\ref{eq:colbf}).
A process is chosen randomly, the same procedure is used as in the case of the
i-packets: the rates are calculated, then a random number is generated and the
same machinery as described in Fig.~\ref{Fig:choice} is processed.

When the collisional excitation \eqref{k_colexc} or ionization \eqref{k_colion}
is chosen, the k-packet transforms to an i-packet. A corresponding transition
(or ion) must be chosen, because i-packet has to contain information, which
element  (ion, level) will be set during the initialization.

Both free-free \eqref{k_ff} and free-bound \eqref{k_fb} transitions change
k-packets to r-packets. A new frequency $\nucmf$ of this packet is in the
free-free case determined from the equation
\begin{equation}
 \label{Eq:ff_emission}
 \int\limits_{\nu^\cmf}^\infty\,\text{d}\nu
 \left(\eta_{j_{\text{I}_j}}^\text{ff}(\nu)\right)^\cmf =
 z\int\limits_{0}^\infty\,\text{d}\nu
 \left(\eta_{j_{\text{I}_j}}^\text{ff}(\nu)\right)^\cmf,
\end{equation}
$z\in[0,1]$ is chosen randomly. In the case of free-bound transition the equation
is as the same as the Eq.~(\ref{Eq:bffreq}).


\section{The general code structure} \label{sec:codeStruct}
The current version of the code calculates radiative transfer through a given
medium. The code can handle a broad variety of media: stellar atmosphere,
circumstellar disc, expanding atmosphere of supernova, etc. The main advantage
of using the MC code is that we solve the problem in {\trid} (or how many
dimensions we want). Thus we are not restricted by the requirement of the model
symmetry and radiative transfer through an atmosphere with irregular mass
distribution can be calculated. On the other hand, the MC method is difficult
to apply to optically very thick media. The calculation is extremely time
expensive because the  number of interactions of every photon  (in this case
packet) is very huge. This is the case of deep layers of the stellar
photosphere. In our stellar wind case this problem does not appear.

\begin{figure}
 \centering
 \includegraphics[width=0.3\textwidth]{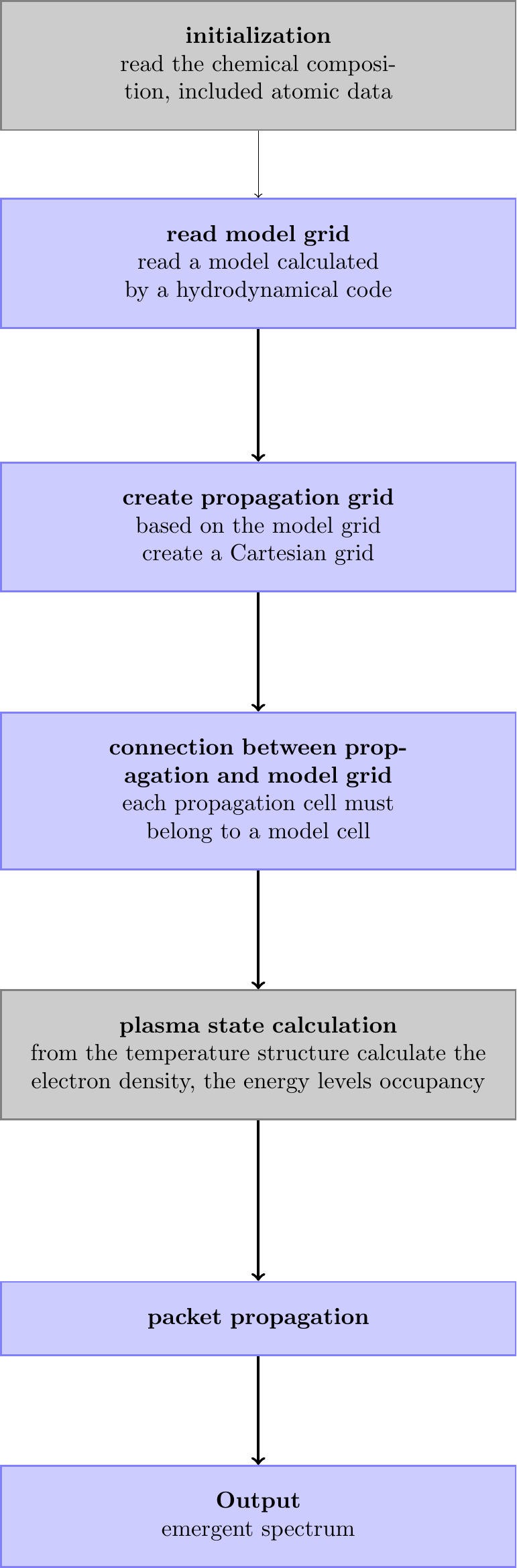}
 \caption{Scheme describing the actual code routine.}
 \label{Fig:flow}
\end{figure}

The general code structure is in the flow chart~Fig. \ref{Fig:flow}. The main
input of the code is a precalculated model of the structure of the flow. This
part of the input includes density, temperature, and velocity profiles. The
next part of the input includes information about the star: effective
temperature, luminosity, stellar radius, and a lower boundary condition, which
can be a photospheric spectrum calculated by another stellar atmosphere code.
The main output of the Monte-Carlo code is the emergent flux.

The code currently assumes the local thermodynamical equilibrium, but it is
possible to use precalculated populations of energy levels from another code
(e.g. \PoWR, \citealt{Hamann2003, Hamann2004}, \citealt{Sander+2015}).

There are two types of numerical grids used in our code: the model grid
(hereafter {\modgrid}) and the propagation grid (hereafter {\propgrid}), see
Appendix~\ref{Ap:numgrids}. The {\modgrid} represents the hydrodynamical model
(input) in isolated points. Direct radiative transfer calculations in
{\modgrid} can be difficult and depend on {\modgrid} geometry. To avoid this
property, the {\propgrid} is created. The {\propgrid} is a Cartesian grid,
which allows radiation transfer calculations regardless the input model
geometry. Every {\propgrid} cell is connected to a {\modgrid} cell. We assume
every physical quantity in the {\propgrid} cell to be constant. Creating the
{\propgrid} is done only at the beginning of the calculations. The connection
of {\modgrid} and {\propgrid} is described in the following subsection. All
technical information about {\modgrid} and {\propgrid} is in
Appendix~\ref{Ap:numgrids}.

The basic {\propgrid} is a regular shaped grid. The {\modgrid} can be very
irregular: with a larger number of grid points in some areas and less grid
points otherwise. To avoid unnecessary dense {\propgrid} we developed an
adaptive grid. There are two types of grids: octgrid, see Fig.~\ref{Fig:dg8},
and one-level subgrid, see Fig.~\ref{Fig:dgijk}. Both grids are more deeply
described in the Appendix~\ref{Ap:numgrids}.
\subsection{A connection between the model and the propagation grid}
The basic {\propgrid} is based on the {\modgrid} size and the density of model
grid points. The dimensions of the {\modgrid} are equal to the Cartesian
maximal dimensions of the {\propgrid}, thus the {\modgrid} fits to the rectangular
parallelepiped shape of the \propgrid.

During calculation of propagation of an energy packet through the {\propgrid},
we need actual values of density, temperature, etc., which are stored in the
{\modgrid}. Since both grids are independent, the definition of their
connection is necessary. Every {\propgrid} cell (hereafter PC) must `know'
which {\modgrid} point ({\modgrid} line or {\modgrid} surface for {\dvad} and
{\jednad} models, respectively) it belongs to. The corresponding {\modgrid}
point (line / surface) is chosen using following steps:
\begin{enumerate}\label{en:cellAssoc}
 \item Choose a propagation cell represented by its centre with coordinates
   $(x_0, y_0, z_0)$.
 \item Calculate distance $l$ between this cell (i.e. its centre) and all possible
  {\modgrid}
  points after the expression following
  from the metrics theory \citep[a distance of two sets, see][]{Munkres2000}
  \begin{equation}
   l = \min\left(\sqrt{(x - x_0)^2 + (y - y_0)^2 + (z - z_0)^2}\right),
   \label{Eq:vzdalenost}
  \end{equation}
  where $(x, y, z)$ is a set of position vectors defining the whole set of
  {\modgrid} points. In the \mbox{1-D} spherically symmetric case it is a
  surface of a sphere $\left\{x^2 + y^2 + z^2 = R_i^2\right\}$, where $R_i$ is
  the sphere radius; in the \mbox{2-D} axially symmetric case the set is
  defined by circles parallel to the $xy$ plane $\left\{x^2 + y^2 = R_i^2\,
  \land\, z = Z_i \right\}$, where $R_i$ is a circle radius and $Z_i$ is the
  $z$-coordinate of the circle. In the \mbox{3-D} case the set of position
  vectors is a single point.
 \item Select the {\modgrid} point (line / surface) with the lowest distance $l$.
 \item In the {\jednad} case test whether the {\propgrid} point is in the wind area, i.
  e. its centre radius $r_\text{C}$ satisfies the condition $$R_* < r_\text{C} < R_\infty.$$
\end{enumerate}
The physical variables are constant in the {\propgrid} cell volume except of
the velocity field, which must be continuous. Analytical velocity fields are
used or an interpolation methods must be implemented in the case of discrete
velocity fields.

\subsection{Subtleties of the propagation of a packet through the adaptive grid}

Indexes of all six neighbouring cells are stored for each {\propgrid} cell,
which simplifies treatment of energy packets moving through the \propgrid. This
concept is simple for a regular grid, however, it becomes more complicated in
the case of an adaptive grid. If a packet flies from the current cell to
another one, a problem can occur: the cell division is not regular, thus the
packet can move from one grid level to another. This problem must be solved
more sophistically.

\begin{figure}
 \centering
 \includegraphics[width=200pt]{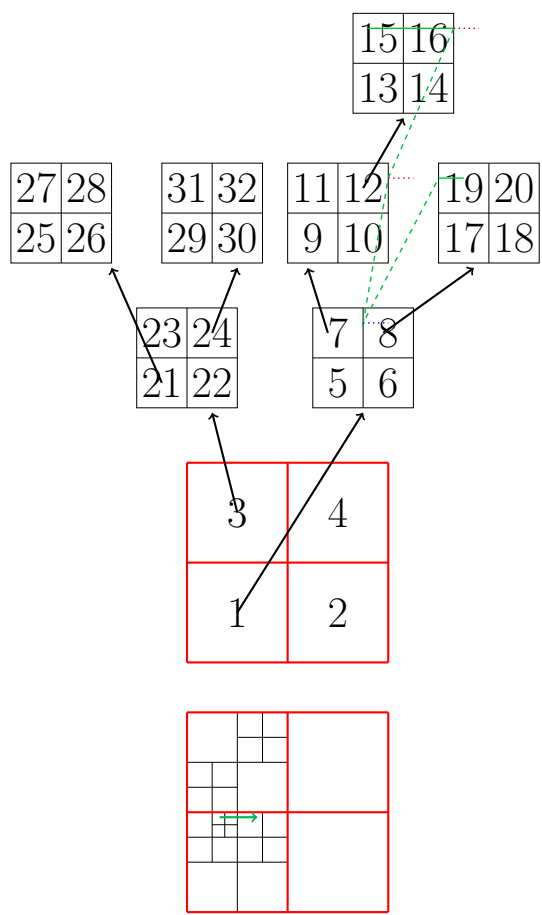}
 \caption{Scheme of the packet propagation through the grid.
  The bottom scheme shows a propagation grid. The upper scheme shows the same grid, but
  expressed in a tree structure. The green line is an example of packet's path. It is
  simple to show this path on the lower graph.
  The upper graph explicitly shows the adaptive propagation grid level structure
  including the illustration of cell indexing. The numbers in the figure describe the
  logic of cell indexing in the code. 
}
  \label{Fig:prop}
 \end{figure}

Every cell `knows' information about its connection to the lower-level cell (a
cell which is divided by the current subgrid) it belongs to, and the first
upper-level cell (the first cell of a subgrid dividing the current cell). These
indexes are zero if the upper- or lower- level grids do not exist. For the
purpose of packet propagation it `knows' also indexes of neighbouring cells of
the same level appertaining to the same lower-level cell. In numbers, if the
index of a neighbouring cell is larger than zero, then the cell exists and it
is at the same level. Otherwise it is equal to zero or $-99$ in the case of the
basic grid, which indicates the end of the computational domain; in the case of
neighbour cell it indicates that the next cell must be found in a different
tree branch, see Fig.~\ref{Fig:prop}. The index zero means that there is no
other cell on the same level, the index -99 is the edge of the whole
propagation grid. The packet propagation through the PC is calculated in the
highest level cells. It cannot continue outside the subgrid but it has to find
the neighbouring cell even in another subgrid.

An example of propagation is shown in Fig.~\ref{Fig:prop}. The packet starts in
the highest level in the cell 15, it moves to the cell 16 because it is the
same subgrid. But there is no cell beyond the current cell and the programme
has to look one level lower -- to the cell 12 which is divided by this subgrid.
But the packet is on the border of this subgrid too, thus once more it goes to
the lower level (cell 7). Here it finds an existing neighbouring cell 8.
Furthermore, it determines an existence of a subcell -- a higher level subgrid.
And it moves to the cell 19, which is calculated using the
Eqs.~(\ref{Eq:indexybunek}) and (\ref{Eq:bunkabaliku}). This cell is the
highest one and the propagation can continue. This process can be generalized
also for subgrids with a variable number of subcells. 

One can see that there are several options of propagation. The most simple case
is the transition from one cell to another within the same subgrid. Then the
Eq.~(\ref{Eq:indexybunek}) can be used. We know the actual cell, the index of
the first cell of the current subgrid. 

If the packet reaches the outer boundary of the current subcell there is no
possibility to propagate in the same level. The programme must skip to the lower
level and continue propagation. Only if the current cell has its subgrid we
have to skip again upwards. The Eq.~(\ref{Eq:indexybunek}) is used for the
index calculation. $N_0$ is saved -- it is a saved index in the lower level
cell. For example the $x$+ border: we know, that the cells on this border have
$N_x = 1$, the indexes $N_y, N_z$ are calculated with the equation
$$N_i = \left\lfloor \frac{r_i-R_i}{w_i}\right\rfloor + 1$$
where $\boldsymbol{R}$ is the coordinate of the $x$- $y$- $z$- corner of the
given subgrid.


\section{Testing the propagation grids}
\label{sec:grid_test}

To check the effectivity of {\propgrid} refinements (construction of the adaptive {\propgrid}) described in
Section~\ref{kap_APG},
we constructed an artificial wind model for the purpose of testing.
Included atomic data were imported from the Opacity project \citep{Delahaye2016}.
The atomic levels were split according to their principal and orbital quantum numbers.
\begin{table}[t]
 \centering
 \caption{Numbers of atomic levels included.}
 \begin{tabular}{p{1cm} c c}
  \hline
  Ion & Max. principal quant. num. &   $N$ of levels \\
  \hline
  \ion{H}{I} & 10 & 55\\
  \ion{H}{II} & 1 & 1\\
  \ion{He}{I} & 6 & 53\\
  \ion{He}{II} & 10 & 55\\
  \ion{He}{III} & 1 & 1 \\
  \hline
 \end{tabular}
\end{table}
We assumed a {\jednad} spherically symmetric wind with a homologous velocity field,
\begin{equation}\label{Eq:homovel}
v(r) = \vinfty \frac{r}{\Rinfty},
\end{equation}
where $\vinfty$ is the terminal wind velocity and $\Rinfty$ is the radius of the outer
model boundary, $r$ is the radial distance. Within the wind we created a clump
represented by a spherical shell with either one or two density peaks.
Then the mass-density of the wind is described by
\begin{equation}\label{test_density}
 \varrho(r) = 
 \begin{cases}
  \densnula \zav{\dfrac{r}{R_*}}^{-2} & R_1\ge r>R_*,\\
  \densnula \zav{\dfrac{r}{R_*}}^{-2} + \densclump(r) &
  R_2 > r \ge R_1,\\
  \densnula \zav{\dfrac{r}{R_*}}^{-2} & \Rinfty>r\ge R_2,\\
  0 & r\ge \Rinfty.
 \end{cases}
\end{equation}
This equation is a power law density distribution in the interval starting at
$R_*$ and ending at $R_\infty$. The clump is represented by an additional
function $\densclump(r)$. For a single-density-peak spherical shell located
between $R_1$ and $R_2$ this function is described by the Gaussian distribution
\begin{equation}\label{test_density2}
 \densclump(r) =
 \densclump\exp \hzav{-\frac{\zav{r-\dfrac{R_1+R_2}{2}}^2}{\zav{R_2-R_1}^2}}.
\end{equation}

The single-density-peak clump with a Gaussian density distribution has the
width $R_2-R_1$ and the density maximum is at $(R_1+R_2)/2$ (the shell centre).
For a double-density-peak clump, the mass-density of the wind is described by
\begin{equation}\label{test_density3}
 \densclump(r) =
 \densclump\hzav{\exp\zav{-\frac{(r-r_{01})^2}{\sigma^2}} + \exp\zav{-\frac{(r-r_{02})^2}{\sigma^2}}}
\end{equation}
with the half-width $\sigma$, and $r_{01}$ and $r_{02}$ are positions of
density-peak centres fulfilling the condition $R_1 < r_{01} < r_{02} < R_2$.
The function $\densclump(r)$ is equal to zero outside the region $(R_1, R_2)$
both for single- and double-density-peak profiles. The quantities $\varrho_0$
and $\densclump$ are density parameters. Numerical values of these parameters
for Eqs.~\eqref{test_density2} and \eqref{test_density3} for our testcase are
listed in the Table~\ref{testclpar}. The distance between {\modgrid} points is
$0.5 R_*$ outside the clump and $0.01 R_*$ inside the clump. 
\begin{table}[t]
 \caption{Parameters used in Eqs.~\eqref{test_density}, \eqref{test_density2} and 
 \eqref{test_density3} as a testcase.}
 \label{testclpar}
 \centering
 \begin{tabular}{p{4cm} c}
  \hline
   effective temperature $T_\text{eff}$ & 14734 K \\
   stellar radius $R_*$ & $16145 R_\odot$\\
   parameter $R_1$ & $6.0 R_*$\\
   parameter $R_2$ & $6.5 R_*$ \\
   outer boundary $\Rinfty$ & $10R_*$\\
   outer boundary velocity $\vinfty$ & $2.875\times 10^9\,\text{cm}\cdot \text{s}^{-1}$\\
   density in an initial time \densnula & $10^{-14}\,\text{g}\cdot \text{cm}^{-3}$\\
   clump density $\densclump$ & $10^{-13}\,\text{g} \cdot \text{cm}^{-3}$\\
   clump inner boundary $r_{01}$ & $6.05 R_*$\\
   clump inner boundary $r_{02}$ & $6.45 R_*$\\
   clump width $\sigma$ & $0.15 R_*$\\
  \hline
 \end{tabular}
\end{table}

Let us introduce a set of variables which describe how well the particular {\propgrid}
(or adaptive {\propgrid}) corresponds to the {\modgrid}. There exists an association
between {\modgrid} and {\propgrid} (see Section \ref{en:cellAssoc}). For each {\modgrid}
cell $I$ we calculate a ratio of the total volume of all {\propgrid} cells associated
with a given {\modgrid} cell and the volume of that particular model cell $V_I$. This
ratio is expressed as
\begin{equation}
 \chi_I = \frac{\displaystyle\sum_{\text{PC}\rightarrow I}V_\text{PC}}{V_{\text{I}}},
\end{equation}
where $V_\text{PC}$ represents the {\propgrid} cell volume belonging to the {\modgrid}
cell $I$ (see Eq.~\eqref{Eq:vzdalenost}, $\text{PC}\rightarrow I$ indexes {\propgrid}
cells belonging to the $I$-th model cell). Ideally, $\chi_I = 1$. However, this condition
is not always satisfied since we try to fit {\modgrid} cells of a general shape by a set
of rectangular {\propgrid} cells. To compare the global quality of connection between the
{\propgrid} and {\modgrid}, we calculated the standard deviation as
\begin{equation}
 \label{meanchidef}
 \sqrt{\left<\chi^2\right>} = \sqrt{\frac{1}{N_\text{\modgrid}}
 \sum_{I=1}^{N_\text{\modgrid}} \left(\chi_I - 1\right)^2},
\end{equation}
where $N_\text{\modgrid}$ is the total number of included model cells connected with at
least one propagation cell. For a good correspondence between {\propgrid} and {\modgrid},
this number should be as close to zero as possible. The relative covering is a percentage
of model cells with at least one {\propgrid} cell connected to.

We compared three types of grids, namely a regular one (Section \ref{basic_PG}), the
octgrid (Section \ref{kap_nested}), and a one-level subgrid (\ref{kap_onelev}). Every
grid type has special characteristics and behaves differently based on the number of
cells.

\subsection{Regular propagation grid test}

\begin{table}
  \caption{Parameters of the regular grids for the single-peak tests.}
  \label{Tab:bgrids}
 \centering
 \begin{tabular}{c c c}
  \hline
   number of cells & $\sqrt{\left<\chi^2\right>}$&
   rel. covering [\%] \\
  \hline
   $5^3$   &  5.77 & 8.2 \\
   $10^3$  &  3.86 & 15.3 \\
   $20^3$  &  4.07 & 40.0 \\
   $50^3$  &  1.29 & 63.5 \\
   $100^3$ &  0.33 & 100.0 \\
   $150^3$ &  0.15 & 100.0 \\
  \hline
 \end{tabular}
\end{table}
\begin{figure}
 \centering
 \includegraphics[width=.5\textwidth]{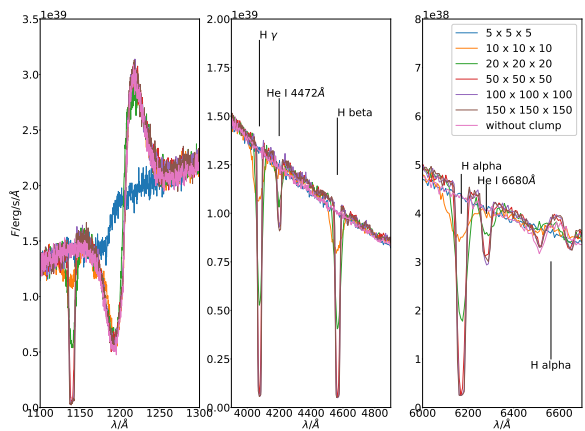}
 \caption{Comparison of spectral line profiles calculated with different
 basic
 {\propgrid}s
in the case of the single clump model.
{\em Left panel:} hydrogen {\Lalpha} line with a P-Cyg profile and the clump
absorption component, {\em central and right panel:} Hydrogen Balmer
lines and neutral helium absorption lines.
Line labels correspond to their vacuum wavelengths.
}
 \label{Fig:sphBasicSpectra}
\end{figure}

First we tested the regular grid. We calculated several spectra with different
division of the basic grid ($\Nbasic$, $\Nbasic = 5^3, 10^3, 20^3, 50^3, 100^3,
150^3$) and plotted the calculated spectra in the region around the hydrogen
{\Lalpha} line (see the left panel of Fig.~\ref{Fig:sphBasicSpectra} and
Table~\ref{Tab:bgrids}). Fig. \ref{Fig:sphBasicSpectra} shows a P-Cygni type
{\Lalpha} line with an additional absorption component caused by the spherical
density clump. The position of this absorption peak at 1128\,{\AA} corresponds
to the velocity of the clump ($0.072 c$). It is nicely seen how the grid
spatial resolution affects the resulting line profiles. For the coarsest
resolution with $\Nbasic=5^3$ the $\Lalpha$ line is very weak, and the clump
absorption is almost not seen. For $\Nbasic=10^3$ the P-Cygni line appears,
however, the clump absorption peak is rather weak. It is stronger for
$\Nbasic=20^3$. For finer grids we obtain better resolution. This is consistent
with the parameter describing the relative covering, which is below 100\% for
low number of {\propgrid} points (see the last column in
Table~\ref{Tab:bgrids}). Finally, the profiles for $\Nbasic=100^3$ and
$\Nbasic=150^3$, are almost identical, which indicates that the grid with
$\Nbasic=100^3$ is sufficient to resolve of the spherical clump. The center and
right panel in Fig.~\ref{Fig:sphBasicSpectra} include several blue-shifted
absorption lines (hydrogen Balmer series and neutral helium lines), which
originate from absorption in the spherical clump. The pure absorption lines
with an exception of a weak {\Halpha} line are not present in the case with no
clump (pink line). The {\Lalpha} and {\Halpha} lines have P-Cygni profiles
originating in the wind.

The calculated $\langle\chi^2\rangle$ for given selected cases is shown in
Table \ref{Tab:bgrids}. The basic (regular) {\propgrid}s (with $10^3$ and
$50^3$ cells in Table~\ref{Tab:bgrids}) do not correspond to the {\modgrid}
well. Increasing the number of {\propgrid} cells to $100^3$ or $150^3$ fits the
{\modgrid} better (see Table~\ref{Tab:bgrids}). The space diagonal of the cell
is equal $0.87(R_2-R_1)$ or $0.56(R_2-R_1)$ ($R_1$, $R_2$ are
introduced in Eq.~\eqref{test_density2}) in the case of the {\propgrid} $100^3$
or $150^3$, respectively. However, this improvement is done for the price of
increasing the number of grid points also in the parts of the model, where it
is not necessary.
\subsection{Adaptive propagation grids tests}
\begin{table*}
 \caption{Computational statistic of the propagation grids in several cases. We show the
  number of cells in the basic grid (second column), the total number of subcells (if
  subgrids are present, third column), the relative CPU time needed for grid creation
  (fourth column) and the relative CPU time of the packet propagation (fifth column). The
  reference values are corresponding CPU times needed to create the regular grid $150^3$.
  For each grid the standard deviation $\langle\chi^2\rangle$ Eq.~\eqref{meanchidef} is shown.
  This quantity describes the quality of the {\modgrid} cell covering.}
 \centering
 \begin{tabular}{c c c c c c c}
  \hline
   num. of cells ({\Nbasic})& num. of
   subcells & num. of virtual points& 
   \multicolumn{2}{c}{time of calculation} &
   $\left<\chi^2\right>$ & rel. covering\\
   (basic grid) & & & creation 
   & pack. prop. & & \%\\
   & & & relative to $t_1$ & relative to $t_2$\\
   & & & $t_1 = 0.617\,$s & $t_2=267.18\,$s\\
  \hline
   \multicolumn{3}{l}{regular, see Fig.~\ref{Fig:dbleClumpSpectra1}}   & & & \\
             $10^3$         & --      & --      & 0.001  & 0.150   & 3.87 & 15.3  \\
             $50^3$         & --      & --      & 0.1   & 0.380   & 1.03 & 63.5  \\
             $100^3$        & --      & --      & 0.7   & 0.680   & 0.33 & 100.0 \\
	     $150^3$        & --      & --      & 1    & 1.00    & 0.15 & 100.0 \\                                        
   \multicolumn{3}{l}{octgrid}        & & & \\
             $10^3$         & 3760    & $10^3$ & 0.06   & 0.178   & 1.97 & 92.9 \\
             $10^3$         & 40768   & $10^4$ & 0.3   & 0.256   & 0.61 & 100 \\
             $10^3$         & 404788  & $10^5$ & 3   & 0.452   & 0.11 & 100 \\
             $10^3$         & 4223624 & $10^6$ & 231 & 0.908   & 0.06 & 100 \\
             $50^3$         & 424     & $10^3$ & 1   & 0.396   & 1.02 & 100\\
             $50^3$         & 18496   & $10^4$ & 6   & 0.401   & 0.68 & 100\\
             $50^3$         & 328720  & $10^5$ & 60  & 0.484   & 0.17 & 100\\
             $50^3$         & 899904  & $10^6$ & 770 & 0.897   & 0.06 & 100\\
   \multicolumn{3}{l}{one-level subgrid}        & & & \\                                                
             $10^3$         & 965     & $10^3$ & 0.03   & 0.174    & 2.23 & 100 \\
             $10^3$         & 45956   & $10^4$ & 0.2   & 0.289    & 0.45 & 100 \\
             $10^3$         & 1631479 & $10^5$ & 10   & 0.809    & 0.07 & 100 \\
             $10^3$         & 54463204& $10^6$ & 483   & 2.426    & 0.06 & 100 \\
             $50^3$         & 0       & $10^3$ & 0.5   & 0.393    & 1.30 & 63.5\\
             $50^3$         & 1400    & $10^4$ & 5    & 0.403    & 0.98 & 100 \\
             $50^3$         & 149804  & $10^5$ & 47   & 0.443    & 0.15 & 100 \\
             $50^3$         & 6357511 & $10^6$ & 3767  & 1.002    & 0.04 & 100 \\
             $100^3$        & 0       & $10^3$ & 4    & 0.749    & 0.33 & 100 \\
             $100^3$        & 8       & $10^4$ & 38   & 0.750    & 0.33 & 100 \\
             $100^3$        & 27902   & $10^5$ & 402  & 0.767    & 0.29 & 100 \\
             $100^3$        & 1907557 & $10^6$ & 3431 & 0.830    & 0.05 & 100 \\
  \hline
 \end{tabular}
 \label{Tab:grids}
\end{table*}
A significantly better and computationally cheaper fit to the model grid can be
obtained using adaptive grids. We tested both methods described in Appendix
\ref{kap_APG} for subgrid creation.

\begin{figure}
 \centering
 \includegraphics[width=.5\textwidth]{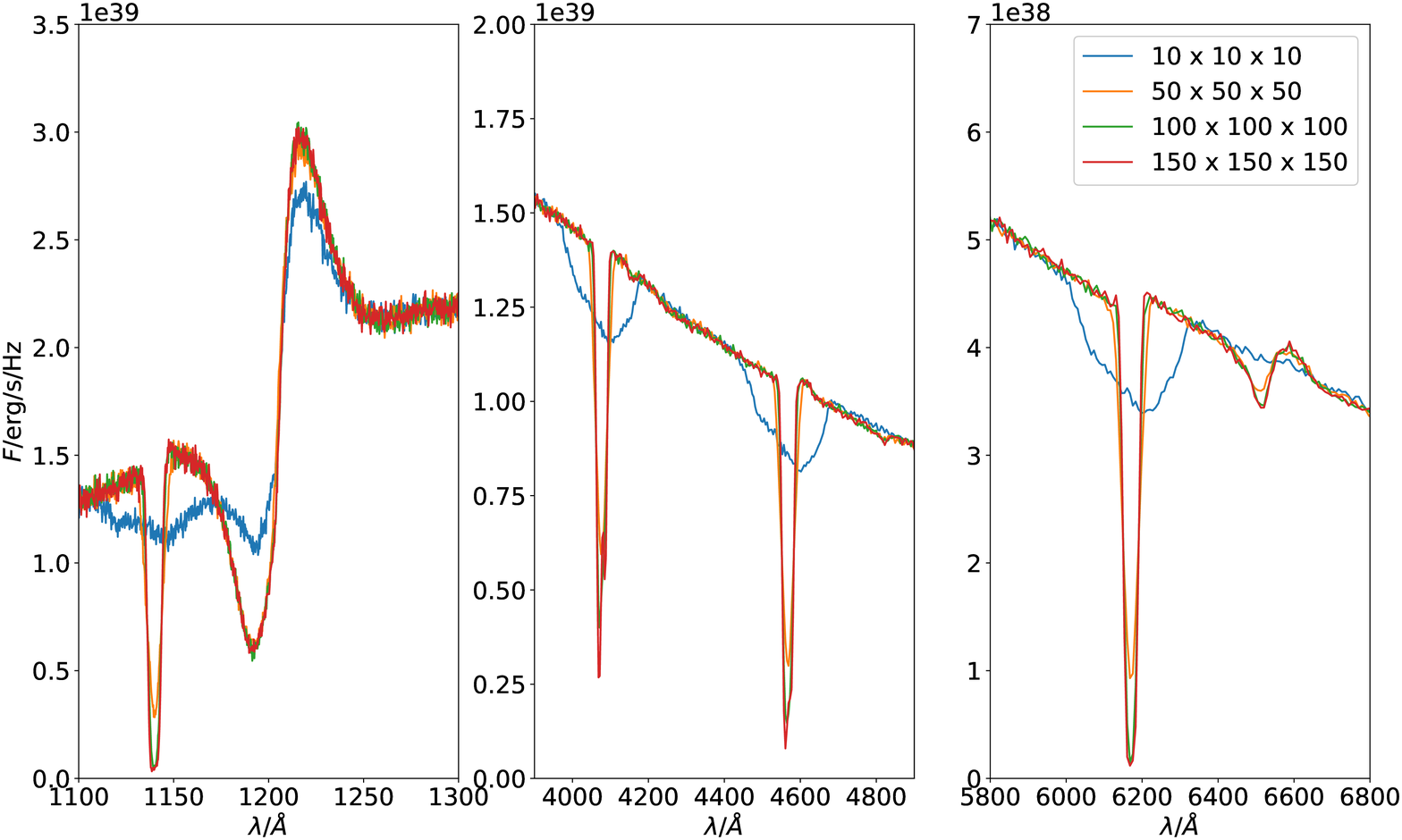}
 \includegraphics[width=.5\textwidth]{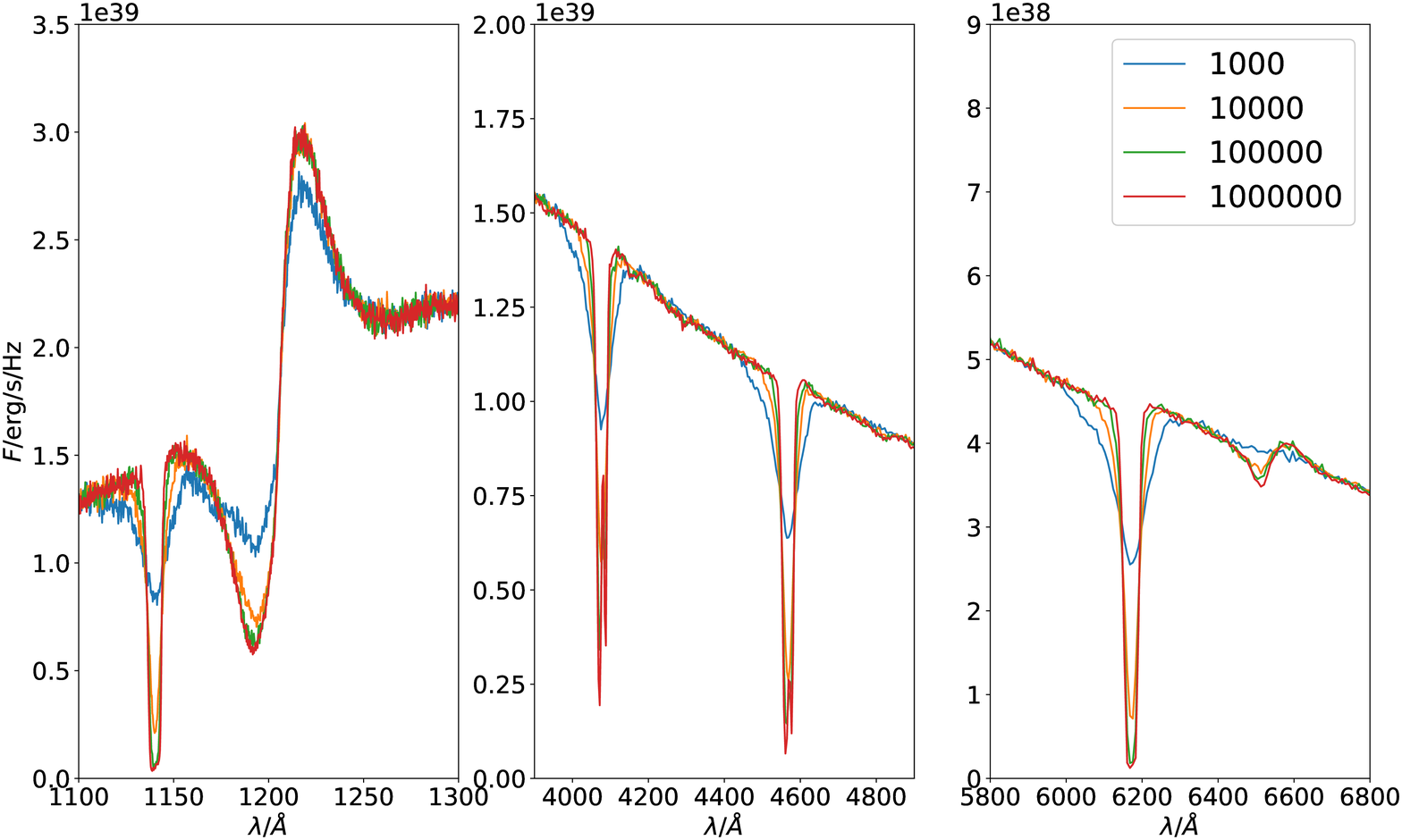}
 \includegraphics[width=.5\textwidth]{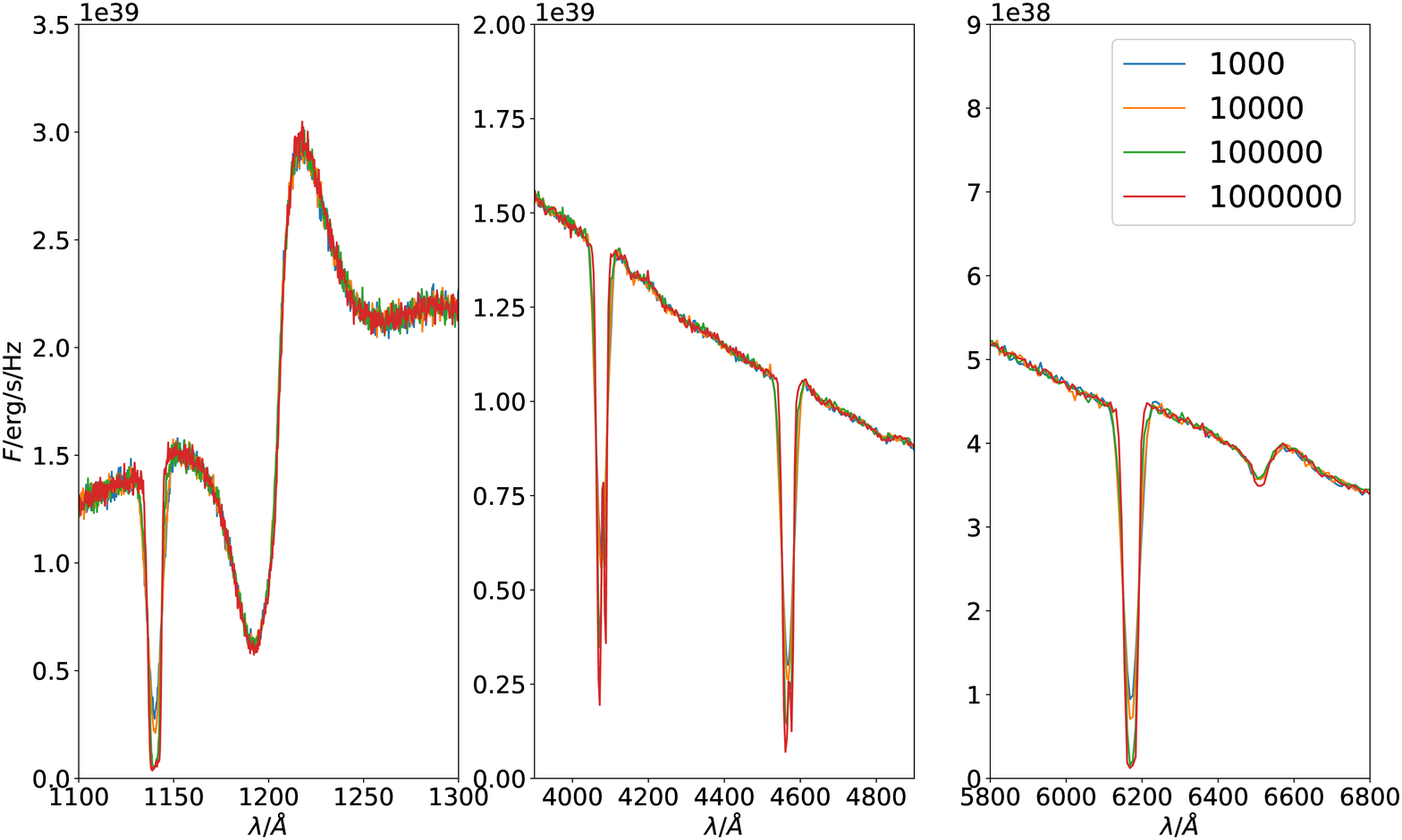}
 \caption{Comparison of spectra calculated with different {\propgrid}s in
 the case of the double clump density profile.
 \emph{Upper panels:} with different number of cells in the basic \propgrid. 
 \emph{Middle panels:} the octgrid: with different number of the virtual
  points, basic grid size is $10\times 10\times 10$.
 \emph{Lower panels:} the octgrid: with different number of virtual points, basic grid size is 
  $50\times 50\times 50$.
  }
 \label{Fig:dbleClumpSpectra1}
\end{figure}

\begin{figure}
 \centering
 \includegraphics[width=.5\textwidth]{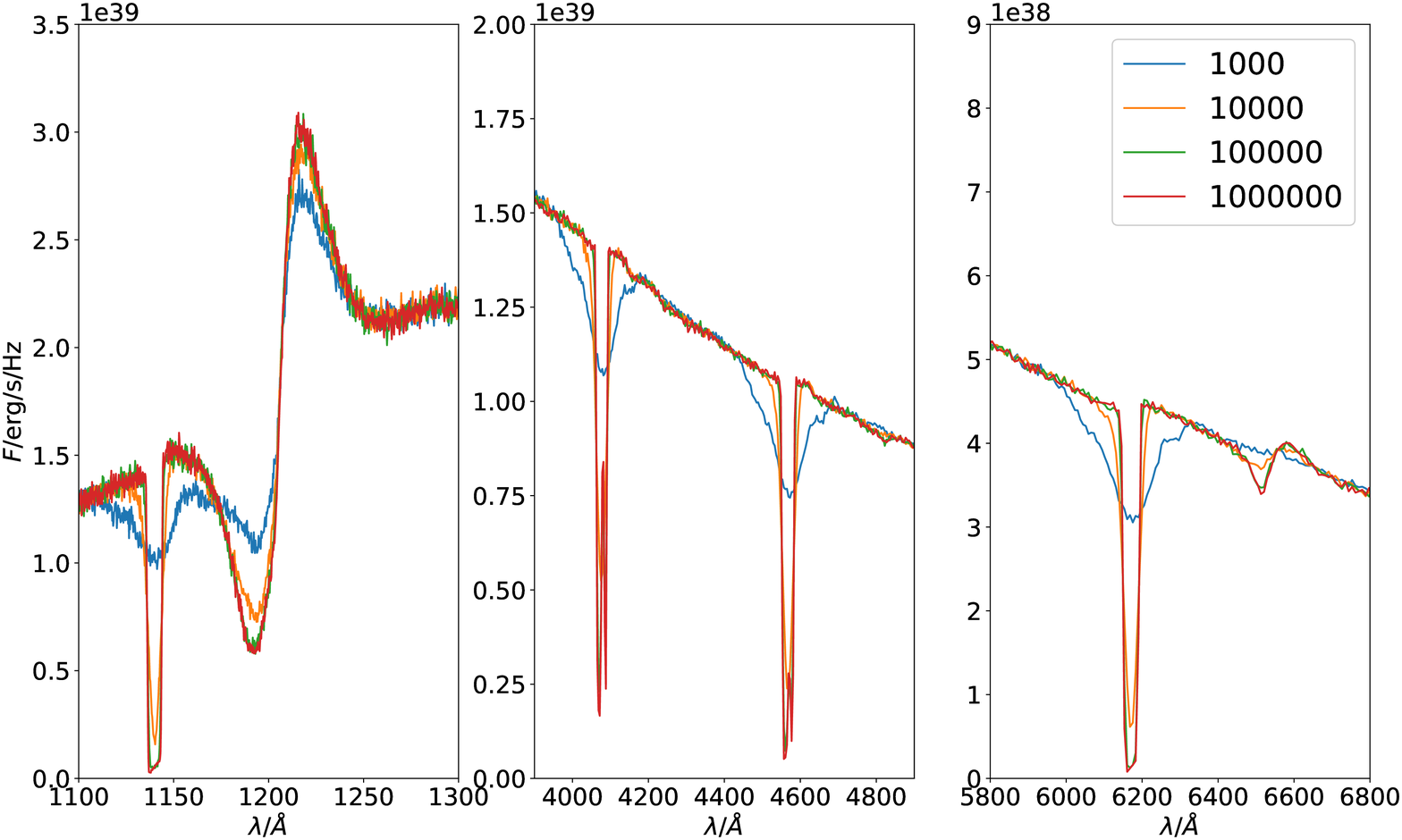}
 \includegraphics[width=.5\textwidth]{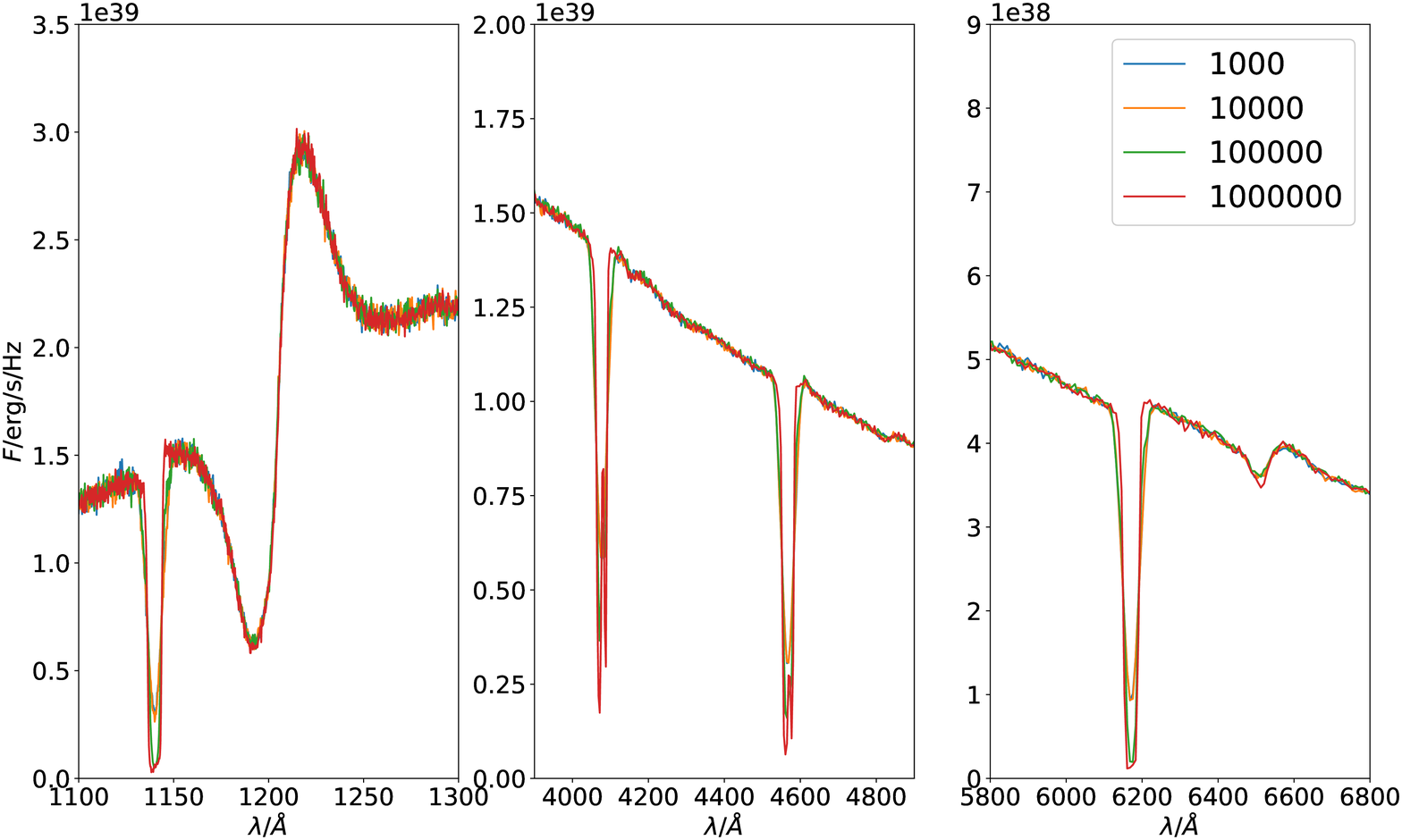}
 \includegraphics[width=.5\textwidth]{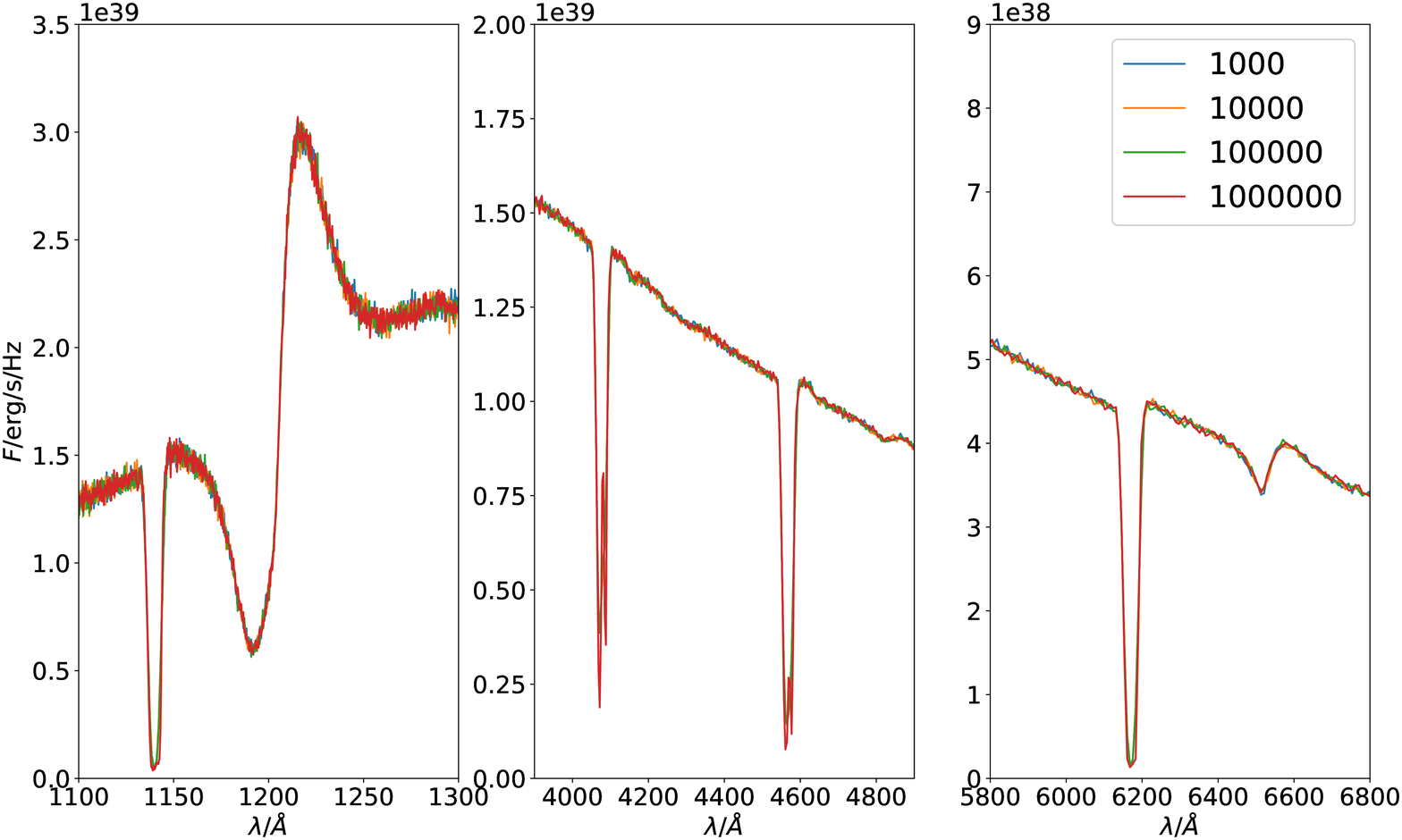}
 \caption{
 Comparison of spectra calculated with different {\propgrid}s in the case
 of the double peak density profile.
  \emph{Upper panels:} the one-level type of the propagation
  grid with the number of the virtual points as a parameter, $10\times 10\times
  10$ basic grid.
  \emph{Middle panels:} one-level grid type, $50\times 50\times 50$ basic grid.
  \emph{Lower panels:} one-level grid type, $100\times 100\times 100$ basic grid.
  }
 \label{Fig:dbleClumpSpectra2}
\end{figure}

\begin{figure*}
 \centering
 \includegraphics[width=.45\textwidth]{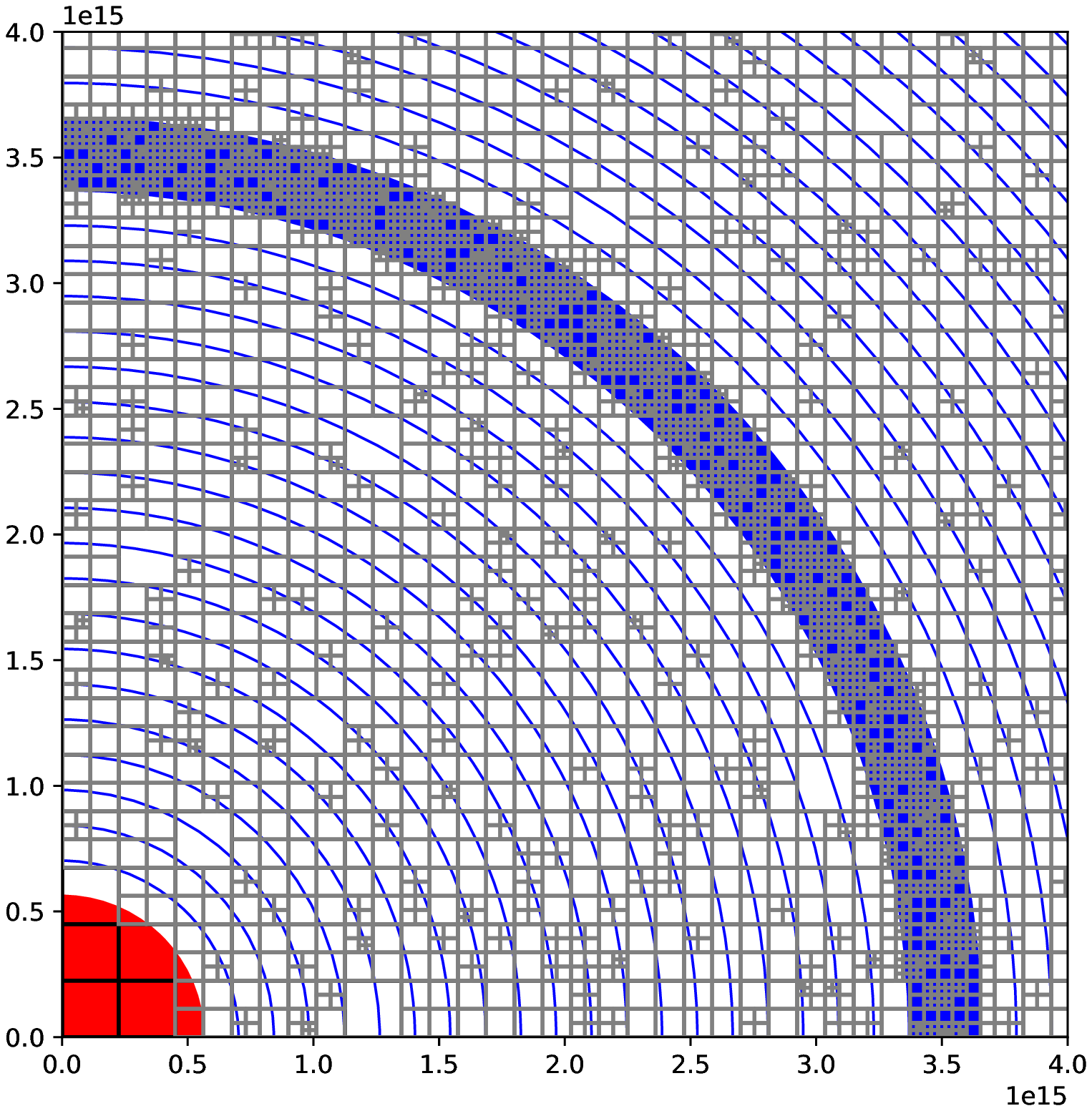}
 \includegraphics[width=.42\textwidth]{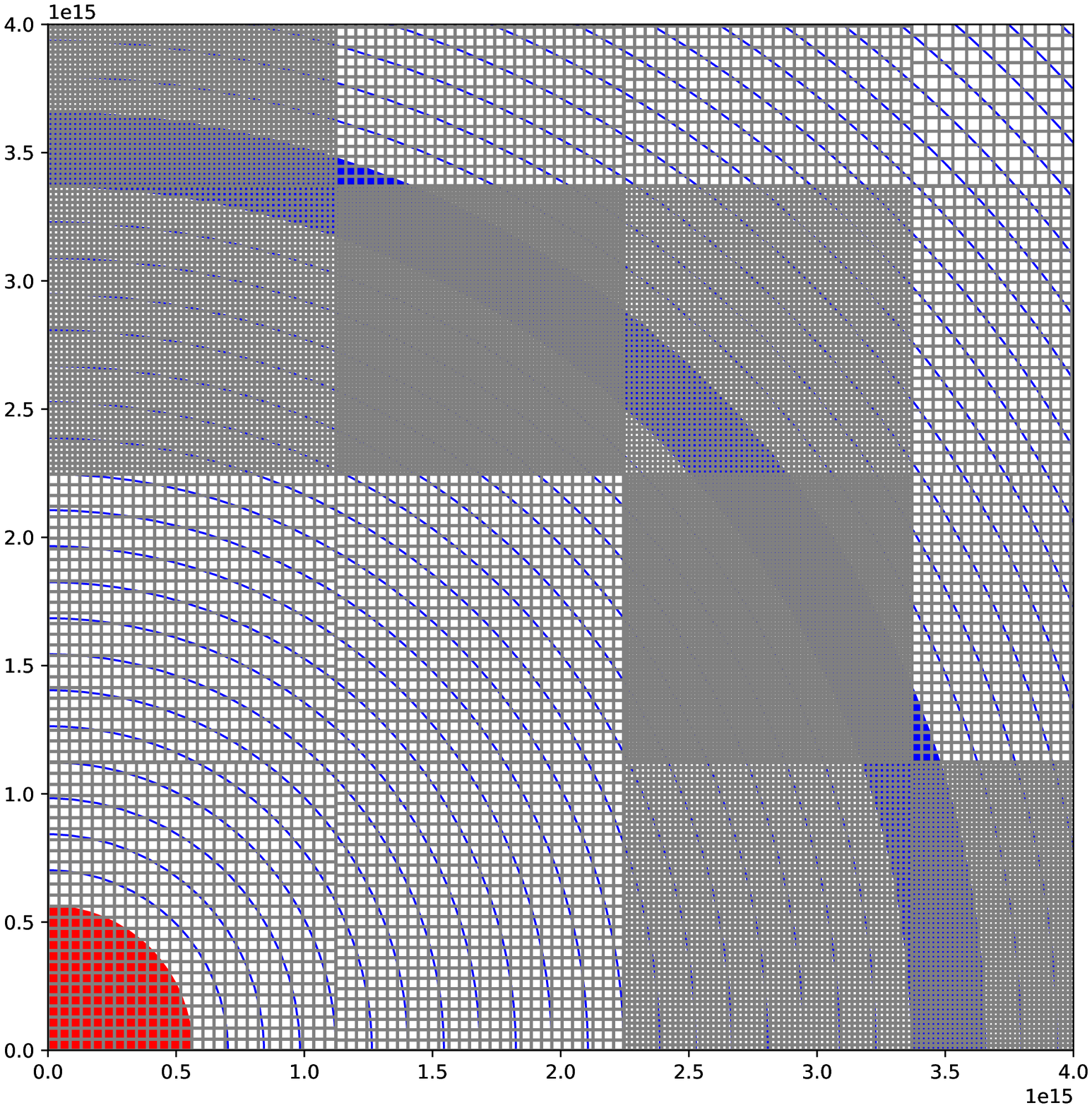}
 \caption{Adaptive propagation grids calculated for the double-clump spherically
  symmetrical model. We have plotted only one quadrant in the zero plane. The red
  ``circle'' is the central star, blue circles represent the model grid shells and the
  grey and the blue Cartesian grid represent propagation grid. It is clear that in the
  area of the clump the density of the {\propgrid} is much larger. \emph{Left:} Octgrid,
  basic grid: $50^3$, $10^6$ virtual points, \emph{right:} one-level subgrid, basic grid:
  $10^3$, $10^6$ virtual points.}
 \label{Fig:sphAdGrid}
\end{figure*}

For the test of the  octgrid we selected basic grids with $10^3$ and $50^3$ cells, both
of them having insufficient relative covering. We created four {\propgrid}s with a
different number of subcells. As a tool for subcell generation we used virtual
points (see Appendix \ref{kap_APG}). The more virtual points the more subcells are generated. We generated $10^3$,
$10^4$, $10^5$, and $10^6$ virtual points. Corresponding numbers of generated subcells
are listed in Table~\ref{Tab:grids}. The numbers in last two columns of the Table
characterize the quality of fit of the {\propgrid} to the {\modgrid}. With the exception
of the $10^3$ basic grid with $10^3$ virtual points all {\propgrid} show 100\%
covering.
For basic $10^3$ {\propgrid} and $10^5$ virtual points we obtain a {\propgrid}
with a standard deviation comparable to the $150^3$ regular \propgrid, which need less
than half of the CPU time for packet propagation. With $10^6$ virtual points we obtain
significantly better standard deviation ($0.06$) with still lower consumption of the CPU
time than for the $150^3$ basic {\propgrid}. The results of calculations with a double
density-peak clump are shown in the Fig.~\ref{Fig:dbleClumpSpectra1} and
\ref{Fig:dbleClumpSpectra2}. One can see that the double profile appears with a better
resolution for {\propgrid} with lower standard deviation. The double minima correspond to
the double shaped density wind structure. We note that insufficient covering of the
{\propgrid} causes erroneous absorption profile with only one peak. Corresponding
adaptive {\propgrid}s are plotted in Fig.~\ref{Fig:sphAdGrid}.

Tests have shown us some advantages and disadvantages of the selected
propagation grid types. The regular grid provides a good approximation,
however, a sufficient resolution requires a dense {\propgrid} in the whole
volume. Much better way is to use the octgrid, which works well for less dense
basic grids: the time necessary to create the octgrid is acceptable since it is
done only once, at the beginning of the calculations, and the time of the
packet propagation is not worse than the regular grid $150\times 150 \times
150$, but with a better resolution with a parameter
$\sqrt{\left<\chi^2\right>}$ equal to 0.06 and compared to the clump
size, the space diagonal (of a rectangular prism) of the smallest {\propgrid}
cell is equal to $0.03\sigma$, where $\sigma$ is defined in
Eq.~\eqref{test_density3}. The octgrid with denser regular grid is not as
efficient as the previous case, grid creating is more time consuming with the
similar efficiency. The one-level subgrid type consumes the most time for the
grid creation and a large amount of memory as well.

The propagation cell input parameters must be chosen wisely. This can be seen in the
Fig.~\ref{Fig:sphAdGrid}. The right figure shows the one-level subgrid based on the
$10\times 10\times 10$ regular propagation grid. The high density of the subgrid is also
located in the large area outside the spherical clump. For the one-level subgrid the
choice of denser basic grid is more appropriate. One can see that in both cases the basic
grid cells are divided even outside the clump, which is caused by the virtual points
distribution. The virtual points are not distributed in the clump only but also among the
other model cells.

The adaptive grid provides a more suitable approach for the {\modgrid} representation. It
does not slow the packet propagation, however its creation is really time consuming in
some cases. This could be prevented by parallelization of the grid creation, which is a
very difficult task, or enable reading already created propagation grid, if the old one
is sufficient enough.

\section{Physical processes in the outflow}
\label{Sec:physProc}
The state of the matter (such as temperature, electron density, etc) must be
determined before the packets propagation calculation. In this section we
introduce the implemented approximations for the plasma state calculation and
the approximation included in the radiative part of the code.

\subsection*{Ionization and excitation state of plasma}
The ionization equilibrium calculation plays a crucial role in radiative
transfer. One of the best ways is to solve the statistical equilibrium
equations. The ionization state and the occupation numbers for the atomic
energy levels depend on the radiation field and the temperature. The
implementation is not easy, thus we temporarily use a simpler approximation of
LTE.

The LTE approximation presupposes the thermodynamic equilibrium for the
ionization fraction and energy levels populations is satisfied locally (on
small distance scales). Thus all occupancy numbers depend only on local values
of temperature and electron density, via the Boltzmann equation
\eqref{Eq:ap:Boltz} and the Saha equation \eqref{Eq:ap:Saha}. These equations
do not provide reasonable approximation of excitation and ionization balance in
hot star winds. The NLTE approximation is more appropriate.

The partition function is also needed to determine plasma state.
Eq.~\eqref{Eq:ap:Part} is used in the code as an initial approximation. Later on more
appropriate approximation will be used.

\section{Computational details}
\label{kapitola_implem}

The energy flow is in the MC method represented via the packet transport. A
packet is a MC energy quantum. The initial packet position is the stellar
surface (the inner boundary condition). A packet propagates through the ejecta
until: it reaches the inner boundary, it reaches outer boundary, or it interacts
$n_\text{max}$ times (a user defined integer). Spectrum is calculated only if
the packet reaches the outer boundary.

Since the propagation of packets is independent it can be parallelized quite
easily. Every processor propagates $n$ packets and the spectrum is calculated
using informations of all propagated packets. The code is parallelized with the
MPI (mpich) library. 
\subsection{r-packets}
Every time the r-packet is propagated the optical depth must be calculated. The
optical depth is composed of line and continuum parts. The line optical depth
is limited in frequency, and, consequently, in differentially expanding media
also in space, by a narrow spectral profile. The continuum optical depth is
nonzero for any frequency.

It is advantageous to create a list of all possible atomic transitions. Every
line has its transition frequency and we arrange the list according to this
frequency. The packet remembers then the last line it visited. The next line
can be only the one closest to the last line. This saves the computational
time.

Opacities in continua are more computationally expensive. The photoionization
cross section depends on the CMF frequency. Since the cross section is not
expressed as an analytical formula but stored as a table for specific values of
frequency, interpolation is necessary. This process consumes plenty of time.
User can choose an approximate way:  choose number of included cross sections
for each element. The number can be larger than zero (number of levels with
included photoionization cross sections, other levels' photoionization cross
section is zero), zero (no photoionization) or $-1$, all possible data included.  
\subsection{i-packets}
The i-packet processing -- calculation of its rates and the inner transitions
can be very time consuming. We have implemented a tool to make the computations
faster. The most time consuming parts of calculation of the i-packet rates are the
recombination parts, see Eq.~\eqref{Eq:ap:bfdown}: the integrals
must be calculated every single step. Integrals contain exponentials which slow
calculations, and, consequently, also the packet propagation down. We prefer to
use the pre-calculated tables. These tables are created during the
initialization processes. The function values are saved for every included ion
with the photoionization cross section data are dependent on temperature.
The intervals are split up linearly, thus the corresponding interval for the
given temperature can be calculated with a simple interpolation. 

\subsection{k-packets}
The main challenge of the k-packets implementation is in the calculation
optimalization. Every time a packet changes into the kinetic energy all
possible rates must be calculated, which takes some time, especially in the
case with many ion levels included.

The first method to save some computation time is to order the rates from the
largest to the smallest one.
The time needed for the calculation which process was chosen gets lower -- in
the code the corresponding rates are saved in an array. The code generates a
random number, which is multiplied with the total rate. Then it goes through
the whole array and looks for an interval corresponding to the randomly
calculated rate. If the first rates are the largest one, the most of the
finding will end in the beginning and do not have to go through all indexes,
which clearly saves the time.

The second method consists of saving the rates from one last model cell. If the
packet returns to the previous model cell it is not necessary to calculate
rates again. 

\subsection{{\propgrid}resolution and number of packets}
Our code contains many changeable parameters which are important for proper
spectra calculation. This implies that inappropriately chosen parameters may
cause incorrect output. We tested spectral sensitivity on the {\propgrid}
parameters and number of packets.
\begin{figure*}
 \centering
 \includegraphics[width=\textwidth]{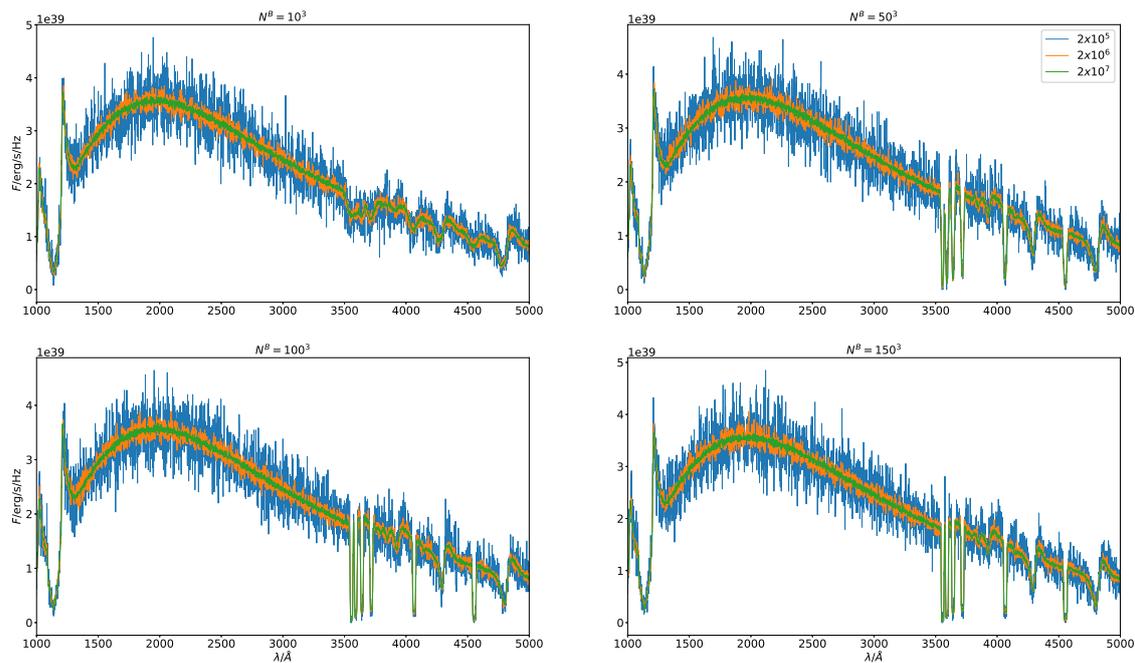}
 \caption{A comparison of calculated spectra for different regular {\propgrid}
 sizes (parameters are written above every figure). Each plot contains a
 comparison of spectra generated from a different number of packets (see the
 legend of upper right plot).}
 \label{Fig:spec_packets}
\end{figure*}
We chose four {\propgrid} sizes $\Nbasic$: $10\times 10\times 10$, $50\times 50
\times 50$, $100\times 100\times 100$, and $150\times 150 \times 150$, and
calculated emergent spectra with a different number of packets. Emergent spectra
are plotted in Fig.~\ref{Fig:spec_packets}. The case of the $10^3$ {\propgrid}
shows insufficient accuracy with too shallow spectral lines, while the other
three grids yield sufficient accuracy. Different tests of the accuracy of the
adaptive {\propgrid}s described in the Section \ref{sec:grid_test} (Figs.
\ref{Fig:sphBasicSpectra}, \ref{Fig:dbleClumpSpectra1}, and
\ref{Fig:dbleClumpSpectra2}) support the fact that the $10^3$ {\propgrid} does
not produce reliable emergent spectrum. What minimum size of {\propgrid} is
good enough depends also on the {\modgrid} resolution. Spectra calculated in
Fig.~\ref{Fig:spec_packets} are well converged in the case of {\propgrid}
$\Nbasic=50^3$, whilst the presence of a spherical clump in
Sec.~\ref{sec:grid_test} needed much better resolution or the adaptive
{\propgrid}. If small structures are present (clumps, etc.) the {\propgrid}
cells must be small enough to describe them.

\begin{figure*}
 \centering
 \includegraphics[width=\textwidth]{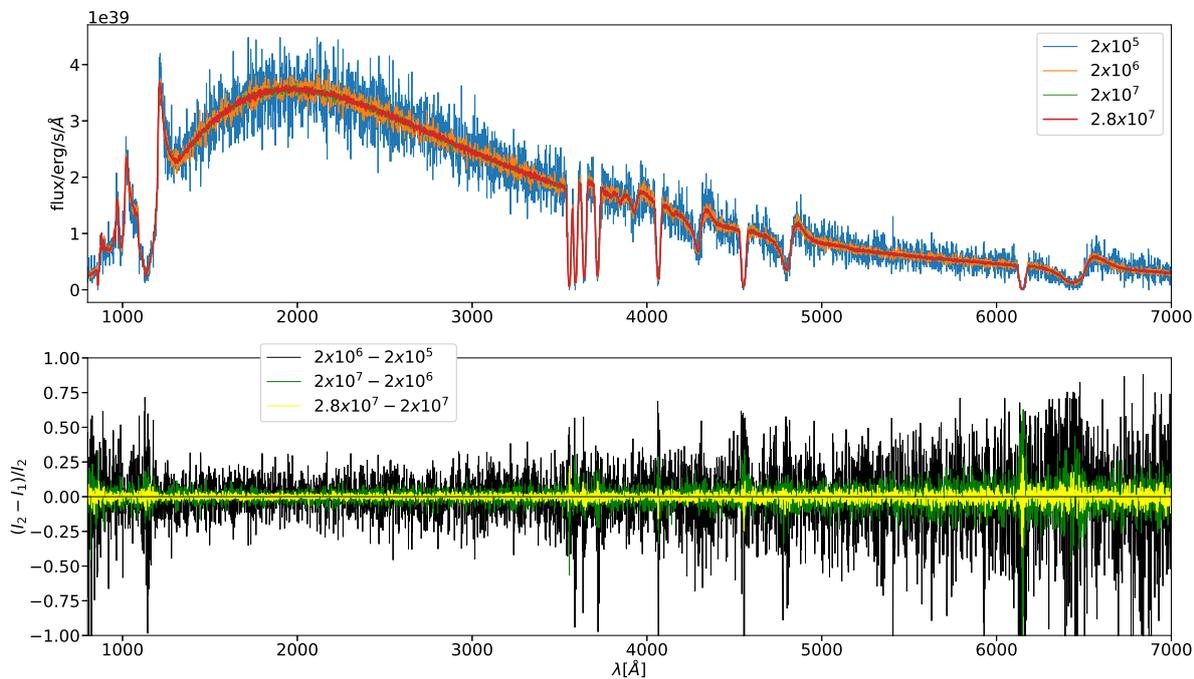}
 \caption{Number of packets is one of the crucial factors for the magnitude of
   ratio: signal to noise in calculated spectra. 
  \emph{Above}: A comparison of calculated spectra for different number of
 packets. \emph{Below}: Relative changes between two spectra for each line.
 Spectra calculated for the grid $\Nbasic=50^3$.}
 \label{Fig:spec_pack_conv}
\end{figure*}

On the other hand, the number of packets affects the noise in the spectra,
which is clearly visible in the plot in Fig.~\ref{Fig:spec_packets}.
Furthermore if a bad {\propgrid} is chosen, increasing the number of packets
does not cause better fit of spectra. Both {\propgrid} and number of packets
must be chosen properly. The relative difference among spectra are shown in
Fig.~\ref{Fig:spec_pack_conv}.
One can see that relative difference between two spectra decreases with
increasing number of packets used for spectra calculation.


\section{A comparison with the {\Tardis} code} \label{Sec:testCase}
To check basic performance of the code and validate it before further steps, we
performed comparison for a specific case using another code, which is somewhat
similar. To this end we chose the wind model as simple as possible, consisting
only from hydrogen and helium because of a low number of spectral lines. The
wind was assumed to be spherically symmetric, radiation flux at the lower wind
boundary was simplified as the Planck function, atomic level populations and
ionization balance were calculated using the LTE approximation. The adopted
velocity field was homologous (Eq.~\ref{Eq:homovel}). Input model parameters
are summarized in Tab.~\ref{Tab:testCase1}. For comparison calculations we used
the code {\Tardis}\footnote{{\url{https://tardis-sn.github.io/tardis/}}}. Our
code reads the same input wind structure (chemical composition, mass density,
velocity structure, temperature) and the atomic data as the {\Tardis} code. 
\subsection{A brief description of the {\Tardis} code}
The {\Tardis} code \citep{Kerzendorf2014} is a Monte Carlo code for radiative
transfer modelling of a supernova explosion. This code is written in Python and
C languages. It allows one to calculate a supernova spectrum for several
parameters describing the physical state of outflow. It can calculate atomic
level populations in both LTE and NLTE approximation, and it can also iterate
the temperature structure. The final spectrum is denoised using the method of
virtual packets. 
\paragraph{Numerical grid}
The {\Tardis} model grid is, similarly to the wind model (see Section
\ref{kap:model_grid}), spherically symmetric and divided to several shells.
Everything is calculated in the spherical coordinates and the code does not use
a propagation grid.
\subsection{Input model for comparison calculations}
There are several parameters which have to be defined. We assumed elemental
abundances to be homogeneous in the whole wind. The mass density dependence in
{\Tardis} is defined by the formula
\begin{equation}
 \varrho(v, t_0, t_\text{expl}) = \varrho_0 \left(\frac{t_0}{t_\text{expl}}\right)
  \left(\frac{v}{v_0}\right)^\kappa,
  \label{Eq:dens_exp}
\end{equation}
where $\varrho_0$ is density in the lowest boundary shell (where $v=v_0$) at an
initial time $t_0$. In this equation $t_\text{expl}$ is a time parameter and
the exponent $\kappa$ has to be chosen by the {\Tardis} user. The
Eq.~\eqref{Eq:dens_exp} is time-dependent, but both {\Tardis} and our code
calculate a single time frame. We selected $t_0 = 1\,\text{day}$,
$t_\text{expl} = 13\,\text{days}$, and $\kappa = -2$. The {\Tardis} code
assumes a homologous velocity field \eqref{Eq:homovel}. The excitation and
ionization equilibrium is calculated using the LTE approximation.

\paragraph{Atomic data}
The {\Tardis} atomic data for hydrogen and helium are sourced from the Kurucz
database. They differ from the atomic data in the Section \ref{sec:grid_test}.
This database contains hydrogen atom with 24 energy levels, neutral helium: 48
energy levels, and once ionized helium: 24 energy levels. We used these data
for our tests.

\subsection{Properties of the calculated models}
The size of computational domain is defined by the velocity field, because of
homologous approximation $v \propto r$.
\begin{table}[t]
 \centering
 \caption{A set of parameters used for the comparison with the {\Tardis}
 code.}
 \begin{tabular}{p{4cm} c}
  \hline
   stellar luminosity & $10^{9.44} L_\odot$\\
   velocity start ($v_0$) & 5000 km/s\\
   velocity end (\vinfty)   & 30000 km/s\\
   inner boundary   &   $8072.4 R_\odot$ \\
   outer boundary   &   $48434.6 R_\odot$ \\
   uniform abundances $\epsilon$ & $\epsilon(\text{H})=0.89$, $\epsilon(\text{He})=0.11$\\
  \hline
   number of packets	& $4\times 10^7$\\
   propagation grid & regular, $N^\text{B} = 150^3$\\
  \hline
 \end{tabular}
  \label{Tab:testCase1}
\end{table}
We calculated models with only line interactions included. Line profiles are approximated
as a $\delta$ function,
\begin{equation}
 \phi_{i}(\nu) \propto \delta(\nu - \nu_i),
\end{equation}
where $i$ denotes the $i-$th line and $\nu_i$ is its transition frequency. The
line
interaction in {\Tardis} can be treated in three modes
\citep[see][Table~1]{Kerzendorf2014}, namely `scatter', where all transitions
are handled as resonance line scattering, `downbranch', where internal
transitions in macroatoms are not considered, and `macroatom', where the full
macroatom scheme is used.

\begin{figure*}
 \includegraphics[width=.5\textwidth]{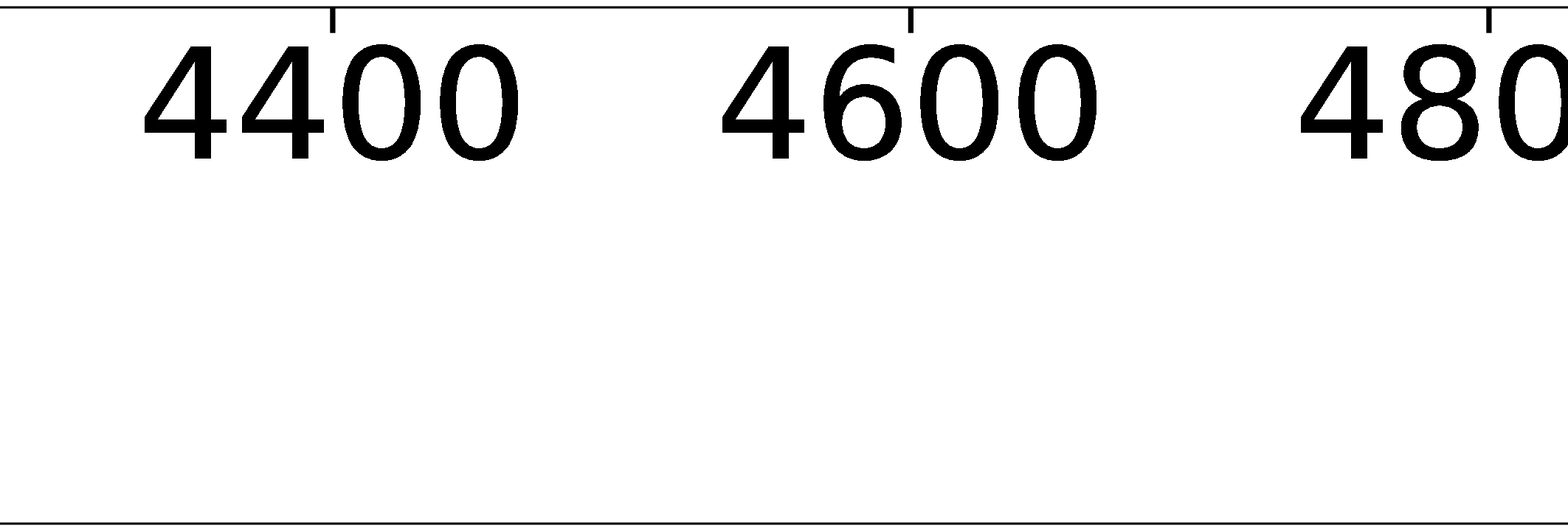}
 \includegraphics[width=.5\textwidth]{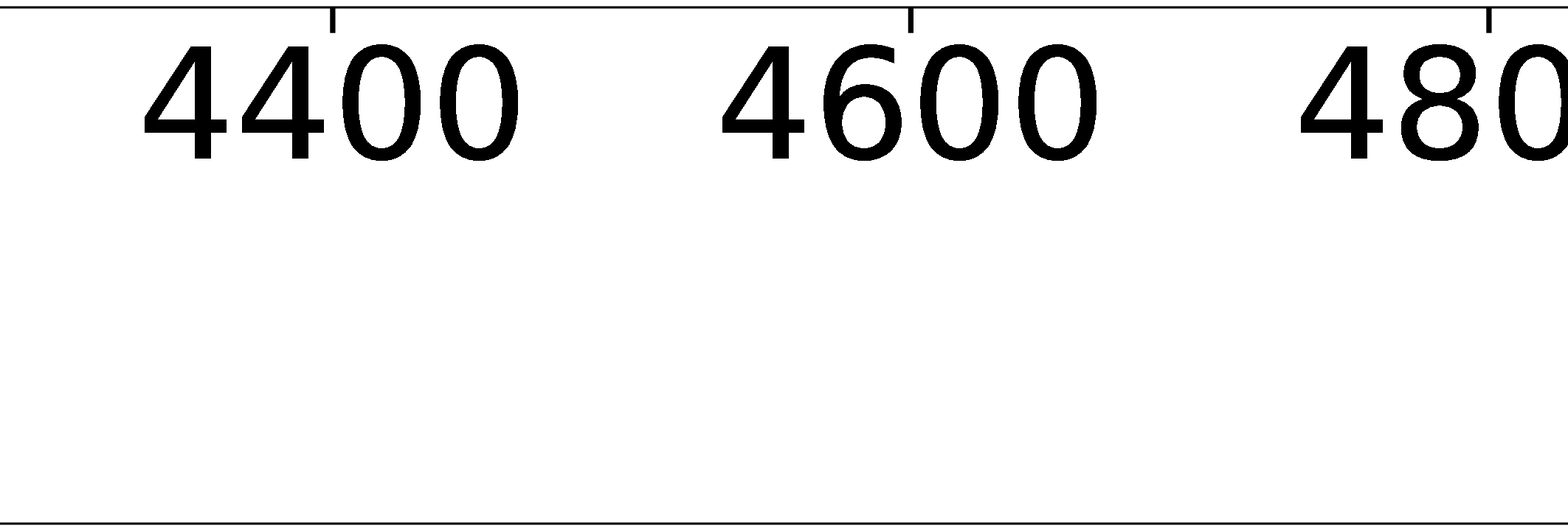}
 \caption{Comparison between emergent spectra calculated by our MC code and
 by {\Tardis}. \emph{Upper panels}: spectra comparison, \emph{Lower panels}: relative difference between
 the spectral fluxes.
 \emph{Left}: Model with only resonance scattering included. Every time the packet
 interacts with an atom, it is directly emitted in the random direction with the same CMF
 frequency. \emph{Right}: Downbranch mode: resonance scattering and fluorescence allowed.
 The deexcitation is possible also to an arbitrary lower atomic level.}
 \label{fig:testcase01}
\end{figure*}
\begin{figure}
 \includegraphics[width=.5\textwidth]{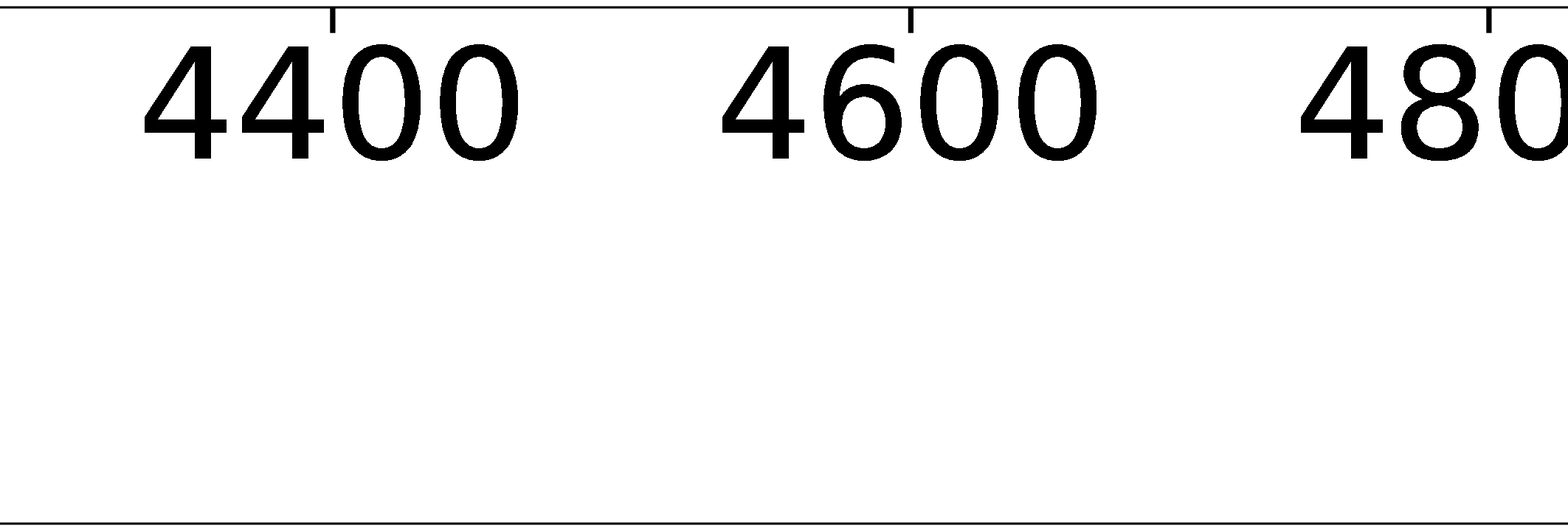}
 \caption{Comparison between emergent spectra calculated by our MC code and
 by {\Tardis}. Line interaction -- the macro-atom mode: the resonance scattering,
 fluorescence, and the internal jumps (in the same ionization state) are included
 in the model.
 \emph{Upper panel}: spectra comparison, \emph{lower panel}: relative difference between
 the spectral fluxes. }
 \label{fig:testcase02}
\end{figure}

The main output is the emergent spectrum averaged over all angles.  We
calculated spectra for all three {\Tardis} line treatment modes (scatter,
downbranch, macroatom) and compared the spectra between the {\Tardis} code and
our code. The results for the resonance and downbranch cases are in
Fig.~\ref{fig:testcase01} and for the macroatom case in
Fig.~\ref{fig:testcase02}.
Strong Lyman and weaker Balmer lines are present in all figures. While
differences in the continuum flux are negligible, differences in lines are more
pronounced. It is seen especially in plots of relative differences (lower
panels of Figs.~\ref{fig:testcase01} and \ref{fig:testcase02}).

There is a difference close to the wavelength corresponding to the wind
boundary velocity $\vinfty$ in the blue wing of the hydrogen {\Lalpha} line. It is
caused by a velocity shift between {\Tardis} and our models and would disappear
if the spectra were corrected for it. The velocity shift causes also a
difference near the centre of the {\Halpha} lines for all three test cases.
Generally, the differences in the {\Halpha} line increase with more complex
line treatment approximation. The largest difference appears in the macroatom
case. There are several possibilities to explain these differences. The
{\Tardis} code and our code are not identical. The differences in emergent
spectra can be caused by differences between grids used in both codes. Our code
creates a Cartesian {\propgrid}, which approximates the {\modgrid} and
`pixelizes' the physical model. The {\Tardis} code calculates only with one
spherical grid corresponding to our {\modgrid}.

To understand these differences better, we did several additional tests to see
the spectra sensitivity on several parameters change. The most sensitive is a
change depending on the model velocity structure. The change of $\vinfty$ and
$\Rinfty$ caused large shifts of the $\Lalpha$ and the other Lyman lines. The
Balmer lines similar to $\Halpha$ were not that sensitive and their shifts were not
as significant as those in the Lyman series. These changes reflect the optical
depth of the outflow in particular line frequencies. We tested the lower
boundary condition: the change of temperature changed the whole continuum and
did not provide any better agreement with the Test case. The stellar radius
change affects the effective temperature as well and did not explain any
difference. The tests including the input supernova model change did not bring
a significant change.

To check the sensitivity of our calculations to a grid resolution, we tested
models calculated with our code for several different grid resolutions. Spectra
do not differ significantly in the cases of regular grid $N^\text{B} > 50$.

We found differences also between the plasma states, the electron densities
differ by $0.7\,\%$ in maximum, and ionization fractions are also different by
about $1\,\%$. This affects the opacity structure and, consequently, the Monte
Carlo rates.


\section{Summary and future work}
\label{Sec:concl}

We presented the first version of our new Monte Carlo code for stellar wind
modelling, which currently performs the calculation of synthetic spectra of
objects with an outflow. The MC method was based on Lucy's Macro-atom model
\citep{Lucy2002, Lucy2003}. The Macro-atom approach is a NLTE radiative
transfer treatment. However, as we concentrated on the radiative transfer
problem, the ionization balance and the level population calculation were
evaluated using the Saha and the Boltzmann equation, because it is much more
simple than the calculation using the statistical equilibrium equations. 

The code starts with reading the input model and creates the model grid
(\modgrid), which contains physical properties of the circumstellar outflow.
Now we use a precalculated hydrodynamical (or pure analytical) model of stellar
outflow. Then it creates a computation domain where the propagation of
radiative energy packets (which describe radiation) is processed. This domain
is called {\propgrid} and is Cartesian. Every packet represents an energy which
propagates and changes its form until it reaches the outer boundary. The inner
and outer boundaries are set up to be consistent with the {\modgrid}.

The {\propgrid} is a Cartesian grid composed of cells of a block shape in
general. The {\propgrid} may be regular or adaptive. The parameters of this
grid can affect the output, thus we tested the spectra calculation based on the
{\propgrid} parameters. The test of the adaptive grid possibilities was done
using an artificial `single clump' and `double clump' density model. The model
without such clump can be calculated efficiently using a regular {\propgrid}.
On the other hand, the model with an added clump requires an enhanced grid cell
density in the clump and a sparser (i.e. unchanged) grid cell density
elsewhere. In the case of the regular {\propgrid}, the grid density is the same
in the whole computation domain. It has to be as dense as the {\modgrid} in the
region of its highest resolution. This construction can be very memory
consuming. Based on this fact we created an adaptive {\propgrid}, which has a
high grid density only in regions with highest {\modgrid} resolution and is
sparse elsewhere.

To test the results of radiative transfer, we compared the synthetic spectra
calculated with our code and with the {\Tardis} code \citep{Kerzendorf2014} for
simple test cases with limited photon-atom interactions. We adopted three kinds
of approximations of line interaction treatment. There are differences between
resulting emergent spectra, increasing with the considered complexity of line
interaction. The reason of this difference is most probably the difference in
treatment of Monte Carlo radiative transfer between our code and {\Tardis}.

The code can solve the radiative transfer through the circumstellar outflow
with an analytically described velocity field. As a next step we shall develop
NLTE ionization and excitation balance in full {\trid} models with a velocity
field defined only in discrete points. This will be reported in future papers.

\begin{acknowledgements}
This research has made use of the NASA’s Astrophysics Data System Abstract Service.
JF was partly supported by the Erasmus+ mobility project.
The Astronomical Institute Ond\v{r}ejov is supported by the project RVO:67985815.

This work was partly supported by a grant GA \v{C}R 22-34467S.

Computational resources were supplied by the project "e-Infrastruktura CZ"
(e-INFRA CZ LM2018140 ) supported by the Ministry of Education, Youth and
Sports of the Czech Republic.
\end{acknowledgements}
\bibliographystyle{aa}
\bibliography{./Literatura}
\appendix
\section{Numerical grids}\label{Ap:numgrids}
\subsection{Model grid}
\label{kap:model_grid}
The model grid (hereafter also {\modgrid}) is a set of points in
$n$-dimensional space ($n\in\{1, 2, 3\}$), where all physical characteristics
of a given circumstellar environment are defined. Selected physical quantities
such as temperature are recalculated during model iterations.

The type of the {\modgrid} corresponds to the assumed problem symmetry. For the
most general {\trid} case the {\modgrid} is simply a set of evenly distributed
isolated grid points. For a {\dvad} model the {\modgrid} is represented by
evenly distributed grid lines and for a {\jednad} model the {\modgrid} is
represented by evenly distributed grid surfaces. For example, spherically
symmetric models are {\jednad} and the grid points are characterized only by
radial distance from the grid centre. Consequently, a grid point is a surface
of a sphere with a given radius. 

The initial physical model may be calculated by another (e.g. hydrodynamic)
code, but also input from previous run of the current code is possible. As the
initial model does not need to follow the rules of our {\modgrid} (since it may
come from a different numerical code), a transformation of the input model to
our {\modgrid} definition is necessary in such case.

\subsection{Basic propagation grid}\label{basic_PG}
The energy packet propagation itself is not calculated in the {\modgrid}
directly. Following \cite{Kromer2009b}, a Cartesian grid called the
`propagation grid' (hereafter \propgrid) is created and all energy packet
propagation calculations are processed in this grid. It is advantageous to
locate the star to the central part of the \propgrid, then the stellar centre
is located at the grid origin.

The main advantage of the Cartesian grid construction is that we employ the
relative simplicity of performing calculations of models with variable
symmetries in a Cartesian grid (which is rectangular) compared to using more
general coordinates (spherical, cylindrical, or even more general ones) to
lower the amount of necessary calculations. In addition, if we want to include
a model with a specific general geometry, we have to modify only the part of
the code which reads the input model and the part which connects the model and
\propgrid. The energy packet propagation itself remains unchanged.

\begin{figure}
 \centering
 \includegraphics[width=.49\textwidth]{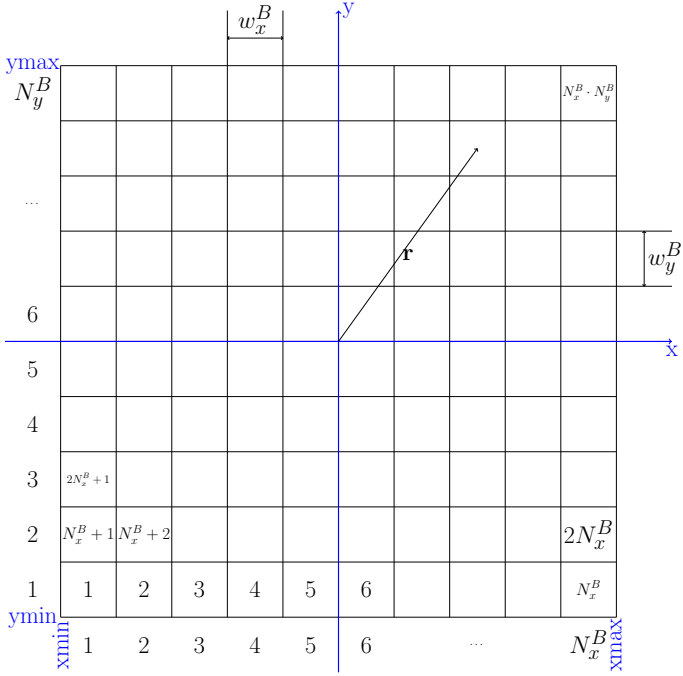}
 \caption{Scheme of the basic {\propgrid} and its coordinates.
   For simplicity, only two dimensions, coordinates $x$ and $y$, are
   plotted here. Each cell has widths $w^\text{B}_x$ and $w^\text{B}_y$.
   Numbers below the figure and on the left denote $n^\text{B}_x$ and
   $n^\text{B}_y$ for particular cells, respectively.
   The cell indexes following from Eq. \eqref{Eq:indexybunek} are written in
   the centers of selected cells.
   Generalization to three dimensions is straightforward.
  }
 \label{Fig:bgcoor}
\end{figure}

\begin{figure}
 \includegraphics[width=200pt]{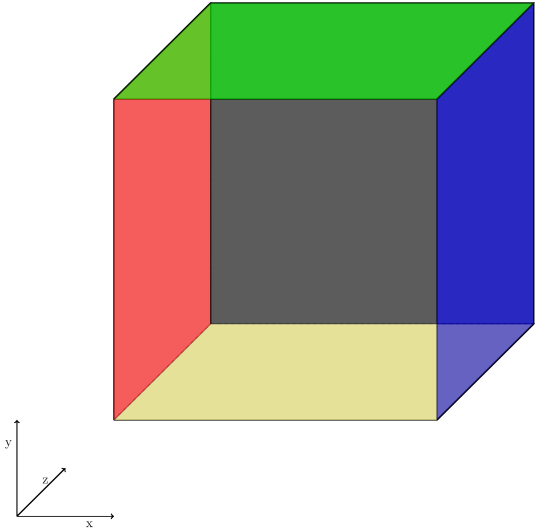}
 \caption{Propagation cell and its orientation to the coordinate system.
  The boundaries are signed as follows: the blue $x+$, the red $x-$,
  the green $y+$, the yellow $y-$, the front $z-$, and the back $z+$.
  The colour cube was created using the code from
  https://tex.stackexchange.com.}
 \label{Fig:bounds}
\end{figure}

The basic {\propgrid} has to be large enough to include the whole {\modgrid}
and consists of $\Nbasic_x \times \Nbasic_y \times \Nbasic_z$ cells, where
$N^\text{B}_i$ represents a number of cells for the coordinate $i$
($i\in\left\{x, y, z\right\}$), see Fig.~\ref{Fig:bgcoor}. All cells have
widths $w^\text{B}_\text{x}, w^\text{B}_\text{y}$, and $w^\text{B}_\text{z}$,
respectively. The widths may be different for each coordinate.
The cell orientation (the plus and minus direction) is in the Fig.~\ref{Fig:bounds}.
The corresponding cell's index
$N(n_\text{x}(x), n_\text{y}(y), n_\text{z}(z))$
is defined as
\begin{multline}\label{Eq:indexybunek}
 N^\textbf{B}(n_\text{x}(x), n_\text{y}(y), n_\text{z}(z)) = \\(N_0 - 1) +
  n_x + (n_y - 1) \cdot \Nbasic_x + (n_z - 1) \cdot \Nbasic_x \cdot
  \Nbasic_y,
\end{multline}
$N^\textbf{B}(n_\text{x}(x), n_\text{y}(y), n_\text{z}(z)) \in \left\{1, \dots,
\Nbasic_x \times \Nbasic_y \times \Nbasic_z\right\}$ (see numbering in
Fig.~\ref{Fig:bgcoor}), $N_0$ is the index of the first cell, in the case of
the basic propagation grid $N_0 = 1$. This equation can be also used for the
determination of neighbouring cells. For example, a neighbour of the cell $N$ in
the $y$-direction placed at $(n_x, n_y + 1, n_z)$ has an index $N + \Nbasic_x$.
The indexes of neighbouring cells are calculated and stored during the
{\propgrid} creation. 

Coordinates of a cell corresponding to an energy packet at a position
described by the radius vector $\vec{r} = (r_x, r_y, r_z)$ can be
calculated as
\begin{equation}\label{Eq:bunkabaliku}
 n_i = \left\lfloor \frac{r_i}{w_i}+\frac{\Nbasic_i}{2} \right\rfloor + 1,
\end{equation}
where $\lfloor\rfloor$ denotes the floor of a number, $w_i$ is the cell width,
and $N_i$ is the number of cells for the coordinate $i$, $i\in\left\{x, y,
z\right\}$. To prevent possible numerical problems in cases when $\vec{r}$
points exactly to the cell boundary, we add a small number $\varepsilon$, which
can be the smallest number allowed by machine accuracy. This helps to overcome
the situation, when the sum $r_i/w_i+\Nbasic_i/2$ should be exactly an integer,
while its numerical representation is slightly lower than the expected value.
Then the expression $\lfloor r_i/w_i+\Nbasic_i/2 + \varepsilon\rfloor + 1$
gives the correct value of the cell coordinate $n_i$.
\subsection{Adaptive propagation grid} \label{kap_APG}
The regular propagation grid is not the most efficient one for every case. As
an example, let us consider a circumstellar disc without a wind. Such disk has
usually a conical shape. To calculate a reliable model we need to choose the
{\propgrid} with small cells, which in the case of a regular grid would be also
outside the disc where very little matter is present. To use memory more
efficiently we need to introduce a propagation grid in which the almost empty
space outside the disk is not considered in such detail.

Every propagation cell represents a homogeneous area of the outflow numerical
model. We want the {\propgrid} to correspond to the {\modgrid} as accurately as
possible. The best and the simplest case is a {\modgrid} with regularly
distributed grid points (every point's location can be represented as
$\mathbf{r} = n_x \mathbf{e}_x + n_y \mathbf{e}_y + n_z \mathbf{e}_z,$ where
$n_{x,y,z}$ were defined in Eq.~(\ref{Eq:indexybunek}) and $\mathbf{e}_{x,y,z}$
is the Cartesian orthonormal basis). In this case, the {\propgrid} can be
created as a regular net with the model points located in the center of each
propagation cell.

In a general case, the {\modgrid} points need not to be distributed equally. In
some regions the number of {\modgrid} points in a unit volume can be much
larger than in the rest of the space. A natural solution handling this problem
is to create a denser {\propgrid} only in the region with more {\modgrid}
points, and not in the whole computational domain. The concept of the adaptive
propagation grid is described, for example, in \cite{Lunttila2012}.

The adaptive propagation grid (hereafter adaptive \propgrid) is created in the
following way: We start from the basic {\propgrid} described in the Section
\ref{basic_PG}. Every grid cell is tested if a subgrid has to be created. For a
{\trid} {\modgrid}, {\modgrid} grid points are counted in every basic
{\propgrid} cell. If their number is larger than one, then a subgrid is
created. Depending on the subgrid type (described later), this process can be
recursively repeated for subgrid cells, if necessary.

Due to an inherent symmetry, {\modgrid} points in {\jednad} and {\dvad} cases
form grid surfaces and grid lines, respectively. Consequently, subgrids can not
be created such as in the {\trid} case, a method for their creation has to be
developed first. We define so-called virtual points for this purpose. The
virtual point is a point with coordinates defined by {\modgrid} surface or line
(2 or 1 coordinates, respectively), supplemented by randomly chosen remaining
coordinates. For example, in the \mbox{1-D} case the radius is defined but the
rest two coordinates, the azimuthal and longitudinal ones, must be calculated. In
total, $N$ virtual points are generated. The number of virtual points
$\mathcal{N}_i$ belonging to the given model cell is proportional to the
spherical shell area
\begin{equation}
 \mathcal{N}_i= N_\text{points}
 \left\lceil\frac{r_i^2}{\sum_{l=1}^{N_\text{MC}}r_l^2}\right\rceil,
\end{equation}
where $N$ is the total number of virtual points, $r_i$ represents the radial
distance of the $i$-th shell and $N_\text{MC}$ the total number of model cells.
Its azimuthal and inclination angle are chosen randomly. Virtual points are
counted for every basic {\propgrid} cell. Similarly to the \mbox{3-D} case, if
this number is larger than one, then a subgrid is created, again with the
possibility of recursively generated subgrids.

The main advantage is that the method of virtual points can be used for a wide
range of geometries without complicated calculations. Virtual points are
created before setting the {\propgrid} up and are not used after the
{\propgrid} is created, thus they do not occupy the computer memory during the
rest of the calculation.

\paragraph{Adaptive {\propgrid} indexing}
The created
subgrid is referred to as a one level higher grid than the basic one. Every
cell of the subgrid has saved the index of its parent cell (lower-level cell)
and the divided cell (the parent cell) in the lower level has saved the first
cell index of the subgrid (upper-level cell). If there is no upper/lower cell,
the corresponding index is equal to zero.

The indexing is illustrated in Fig.~\ref{Fig:prop}. First the basic
grid is numbered as described in the Appendix~\ref{basic_PG}. Then the programme
goes through the grid from the cell with the lowest index and if a cell
contains a subgrid, the subgrid is indexed. In the next step it recursively
checkes the subgrid and looks for its existing subgrids, again starting with
the cells with lowest indexes. The process continues in a loop until it reaches
the end of the given subgrid level. Then it goes back to the lower-level cell and
continues to the cell with the index higher by one.
In following subsections implemented methods of subgrid creation are described.
\subsubsection{The octgrid}
\label{kap_nested}

\begin{figure}
 \centering
 \includegraphics[width=0.49\textwidth]{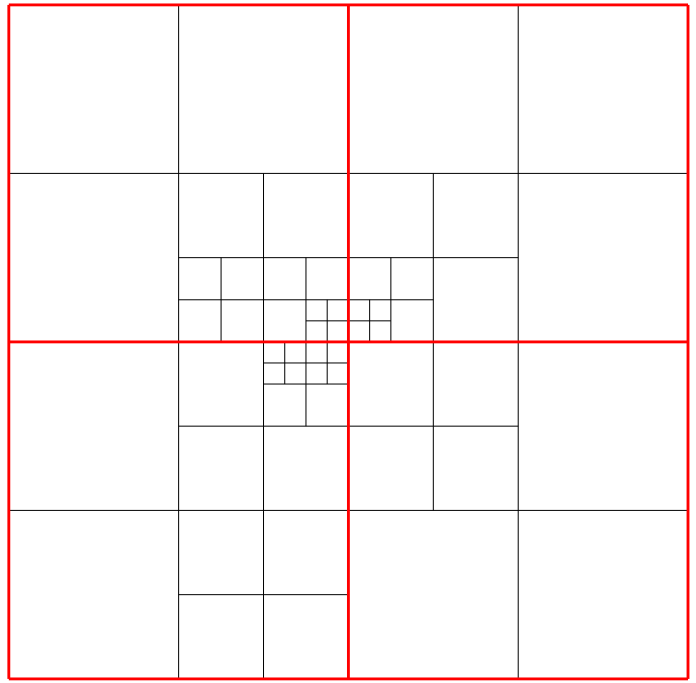}
 \caption{
 Adaptive {\propgrid} 'octgrid'. Every cell can be recursively divided into four (in a \mbox{2-D}
case) or eight (in the \mbox{3-D} case) cells until the predefined
minimum cell limit is reached.}
 \label{Fig:dg8}
\end{figure}

The first subgrid scheme is based on the octree grid. It is called `octgrid'
\citep[e.g.][]{Jonsson_2006, Acreman+2010, Robitaille_2011}. It is created
based on the following rules:
\begin{enumerate}
 \item A cell is divided into $2^3$ subcells, which are in a geometrical sense
  similar with the divided cell. The subcell dimensions are half the dimensions
  of the original cell.
 \item Subcells can be created recursively, the total number of nested cells is
  limited by a minimal cell width.
\end{enumerate}
A~{\dvad} example is shown in the Fig.~\ref{Fig:dg8}. This subgrid scheme is
useful for models with non-regularly distributed grid points as well as for
models with inhomogeneities of model physical parameters or their derivatives.
Such models require larger number of {\propgrid} cells in specific regions and
subgrid nesting allows smooth transition between smallest and largest cells.

This type of grid has been already described in \citet{Barnes1986} and
\citet{Kurosawa_Hillier_2001}. However, while
\citeauthor{Kurosawa_Hillier_2001} used calculation of an emissivity at random
points, we created virtual points.
\subsubsection{One-level subgrid}
\label{kap_onelev}

The second type of grid we call the 'one-level subgrid' (see
Fig.~\ref{Fig:dgijk}). In this case the basic cell is divided into a regular
subgrid, generally with any number of equal subcells. The number of subcells
may differ from the number of the basic grid cells. The grid has the following
properties:
\begin{figure}
 \includegraphics[width=0.49\textwidth]{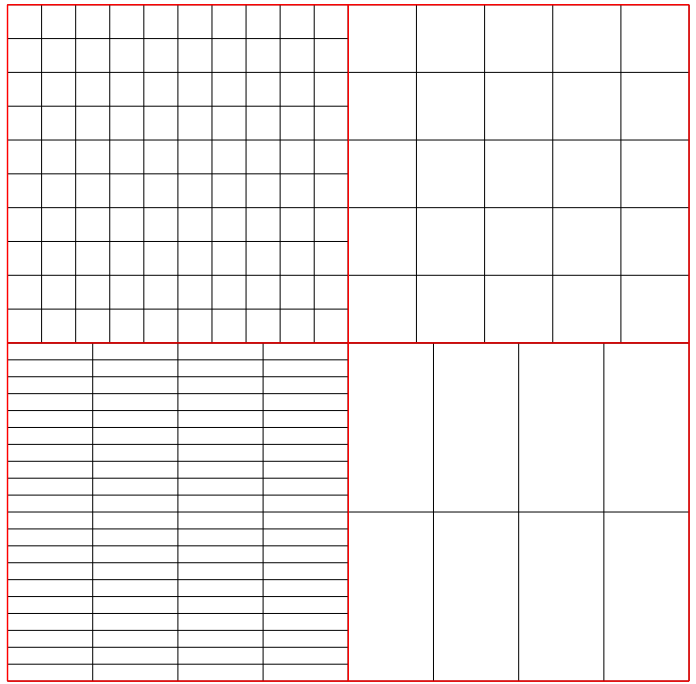}
 \caption{
 Adaptive {\propgrid} one-level subgrid. In every basic cell (red colour) a
 subcell with number of cells $n_x, n_y, n_z$ and variable widths $w_x, w_y$,
 and $w_z$ is created.
}
 \label{Fig:dgijk}
\end{figure}
\begin{enumerate}
 \item A cell is divided into a subgrid $n^\text{S}_\text{x} \times
       n^\text{S}_\text{y} \times n^\text{S}_\text{z}$, where $n^\text{S}_i$ is a
 number of subcells in the coordinate $i$.
 \item Further division of the subgrid is forbidden.
\end{enumerate}

\section{Radiative and collisional rates used in the code}\label{ap:radcolrates}

Here we specify detailed expressions for rates used in our calculations.

\subsection{Radiative rates}
\subsubsection{Bound-bound rates}
The radiative rate for an upward bound-bound transition from the level $i$ to the
level $u$ (excitation) is equal to
\begin{equation}
 R_{iu} = B_{iu}
  \int\,\text{d}\nu\,\phi_{iu}(\nu)J(\nu).
 \label{Eq:ap:bbup}
\end{equation}
The inverse rate (deexcitation from the level $i$ to the level $l$) is
\begin{equation}
 R_{il} = 
   A_{il} +
  B_{il} \int\,\text{d}\nu\, \phi_{il}(\nu)J(\nu).
 \label{Eq:ap:bbdown}
\end{equation}
In these equations, $A_{il}$, $B_{il}$, and $B_{il}$ represent the Einstein coefficients,
$\phi_{iu}(\nu)$ is the line profile of a transition between levels $i$ and $u$, and
$J(\nu)$ the mean radiation field intensity.

For the case of the Sobolev approximation we approximate the line profile by a delta
function,
\begin{equation}
 \phi(\nu) = \delta(\nu - \nu_{iu}),
\end{equation}
where $\nu_{iu}$ denotes the line centre frequency.
Then the absorption rate for transitions from the level $i$ to the level $u$
simplifies to \citep[for derivation see][]{KleinCastor1978}
\begin{equation}
 R^\text{S}_{iu} = \beta_{iu} B_{iu} J_{iu},
 \label{Eq:absSob}
\end{equation}
where the upper index $\text{S}$ stands for Sobolev,
\begin{equation}
J_{iu} = \int \der\nu \cdot\phi(\nu) J(\nu) =
 \int\der\nu \cdot \delta(\nu-\nu_{iu}) J(\nu) = J(\nu_{iu}),
\end{equation}
and
\begin{equation}
\label{eq:betasob_def}
 \beta_{iu} = \frac{1}{\tau_{iu}^\text{s}}
  \left[1-\exp\left(-\tau_{iu}^\text{s}\right)\right]
\end{equation}
denotes a probability that a scattered photon escapes.
The quantity $\tau_{iu}^\text{s}$ is the Sobolev optical depth for the transition
between levels $i$ and $u$.
Similarly, the emission rate from the level $i$ to the level $l$ is
\begin{equation}
 R^\text{S}_{il} = \beta_{il}
  \left(A_{il} + B_{il} J_{il}\right).
 \label{Eq:emiSob}
\end{equation}
where $\beta_{il}$ is defined similarly to the Eq.~\eqref{eq:betasob_def}.

If we treat stimulated emission as negative absorption the absorption and
emission rates become
\begin{equation}\label{Eq:ab-stim:bbup}
 \tilde{R}^\text{S}_{iu} = 
  \left(B_{iu} -
   \frac{n_u}{n_i} B_{ui}\right)
   \beta_{iu}J_{iu},
\end{equation}
and
\begin{equation}\label{Eq:emspont:bbdown}
 \tilde{R}^\text{S}_{il} = 
  \beta_{li} A_{il},
\end{equation}
respectively.

\subsubsection{Bound-free and free-bound rates}
The radiative ionization rate from the level $i$ to the level $p$ (here the
ground level of the next higher ion)
\begin{equation}
 R_{ip} 
  = 4\pi \int\limits_{\nu_{i}}^\infty\,\text{d}\nu
  \frac{\alpha_{ip}(\nu)}{h\nu}J(\nu),
 \label{Eq:ap:bfup}
\end{equation}
and the radiative recombination rate ($i$ is the ground level of an ion here,
the corresponding population is denoted as $n_{i_{l_0}}$), $m$ corresponds to a
level in the next lower ion:
\begin{multline}
 R_{im} = \left(\frac{n_{m}}{n_{i_{l_0}}}\right)^*
  \\ \times 4\pi
  \int\limits_{\nu_{m}}^\infty\,\text{d}\nu
  \frac{\alpha_{mi}(\nu)}{h\nu}
 \zav{\frac{2h\nu^3}{c^2} + J(\nu)}
 \exp\zav{-\frac{h\nu}{\bolk \Telec}},
 \label{Eq:ap:bfdown}
\end{multline}
where $\alpha_{mi}(\nu)$ is the photoionization cross
section from the level $m$ to the level $i$. The equation consists of two terms: spontaneous
($R_{im}^\text{spont}$) and stimulated ($R_{im}^\text{stim}$) recombination rate,
\begin{equation}
 R_{im}^\text{spont}
 = \frac{n_{i_{l_0}}}{n_i} n_\text{e} \phi_{i}(\Telec) 4\pi
 \int\limits_{\nu_{i}}^\infty\,\text{d}\nu
 \frac{\alpha_{mi}(\nu)}{h\nu}\frac{2h\nu^3}{c^2}
 \exp\zav{-\frac{h\nu}{\bolk \Telec}},
 \label{Eq:ap:bfdownspont}
\end{equation}
and
\begin{equation}
 R_{im}^\text{stim} 
 = \frac{n_{m_{\text{L}_0}}}{n_i} n_\text{e} \phi_{i}(\Telec) 4\pi
 \int\limits_{\nu_{i}}^\infty\,\text{d}\nu
 \frac{\alpha_{mi}(\nu)}{h\nu}J(\nu)
 \exp\zav{-\frac{h\nu}{\bolk \Telec}}.
\end{equation}
Here
\begin{equation}\label{eq:sabolf}
 \phi_{i}(\Telec) = \frac{n_{i}}{n^*_{m(0)} n_\text{e}},
\end{equation}
is the Saha-Boltzmann factor.

If we assume that the stimulated recombination is negative photoionization, the
photoionization rate is given by
\begin{equation}
 \label{Eq:photiontilde}
 \tilde{R}_{ip}
 = \gamma_{i} - \frac{n_{p_{\text{L}_0}}}{n_i} n_\text{e}
 \alpha_{i}^\text{stim},
\end{equation}
where
\begin{equation}
 \gamma_{i} -
 \frac{n_{p_{\text{L}_0}}n_\text{e}}{n_{i}} \alpha^\text{stim}_{i}
 = 4\pi \int\limits_{\nu_{i}}^\infty\,\text{d}\nu
 \left(1- \frac{n_{p_{\text{L}_0}}}{n_{i}}
 \frac{n^*_{i}}{n^*_{p_{l_0}}}
 \exp\left(-\frac{h\nu}{k_\text{B}T_\text{e}}\right)\right) J(\nu).
\end{equation}

\subsection{Collisional rates} 
\label{sec:colrat}

For collisional rates from the level $i$ to the level $\anyl$ (which may be $l$, $u$, $m$, or
$p$, see section \ref{Sec:MCRT}) the expressions of $C_{i\anyl}$ for bound-bound and
bound-free transitions are given below. The inverse rates are calculated using $n_i^*
C_{i\anyl} = n_\anyl^* C_{\anyl i}$ \citep[see][Eq.\,9.52]{Hubenyc2015}.
\subsubsection{Bound-bound transitions}

The bound-bound collisional rates are calculated using the \citet{vanRegemorter1962} approximation
\begin{equation}
 C_{iu} = 14.5\cdot n_\text{e}c_0 T^\frac{1}{2}
 \zav{\frac{I_\text{H}}{h\nu_{iu}}} f_{iu}
 \frac{h\nu_{iu}}{\bolk \Telec}
 \exp\zav{-\frac{h\nu_{iu}}{\bolk \Telec}}
 \Gamma\zav{\frac{h\nu_{iu}}{\bolk \Telec}},
 \label{eq:colbb}
\end{equation}
\citep[see also][Eq. 9.58]{Hubenyc2015}, here $c_0 = 5.46510\cdot 10^{-11}$ is
a constant, $I_\text{H}=13.6\text{eV}$ is ionization potential of hydrogen and
$h\nu_{iu}=\epsilon_u - \epsilon_i$ is the transition energy between levels $i$
and $u$.
The function $\Gamma$ is defined as
\begin{equation}
 \Gamma(x) = \max\left(\bar{g},0.276\cdot\exp(x)E_1(x)\right),
\end{equation}
$\bar{g}$ is given by (here $n,n'$ denote principal quantum number, $l, l'$ the orbital quantum
number)
\begin{equation}
 \bar{g}\sim
 \begin{cases}
  0.7, & \text{for transitions} <n,l>\rightarrow<n,l'>,\\
  0.2, & \text{for transitions} <n,l>\rightarrow<n',l'>,\\
 \end{cases}
\end{equation}
and $E_1(x)$ is the exponential integral function. 
\subsubsection{Bound-free transitions}
The collisional bound-free
rates are calculated using an approximate formula
\citep[see][Eq.\,9.60]{Hubenyc2015}
\begin{equation}
 C_{ip} = n_\text{e}\frac{1.55\cdot 10^{13}}{T_\text{e}^\frac{1}{2}}
 \bar{g}_i\frac{\alpha_{ip}}{\frac{h\nu_{i}}{k_\text{B}T_\text{e}}}
 \exp\left(-\frac{h\nu_{i}}{k_\text{B}T_\text{e}}\right),
 \label{eq:colbf}
\end{equation}
where $\bar{g}_i$ is equal to
$$
 \bar{g_i}\sim
 \begin{cases}
  1 & \text{charge of ionization state of level }i\text{ is }0,\\
  2 & \text{charge of ionization state of level }i\text{ is }1,\\
  3 & \text{charge of ionization state of level }i\text{ is} \ge 2.
 \end{cases}
$$
\subsubsection{Free-bound transitions}
 The cross section for the spontaneous free-bound transition from the level $i$
 to the level $m$ is
 \begin{equation}
  \alpha^\text{E, spont}_{i} = \phi_{i}(T_\text{e}) 4\pi
   \int\limits_{\nu_{i}}^\infty\,\text{d}\nu
   \frac{\alpha_{im}(\nu)}{h\nu_{i}}
   \frac{2h\nu^3}{c^2}\exp\left(-\frac{h\nu}{k_\text{B}T_\text{e}}\right)
   \label{ap:fbEspont}
 \end{equation}
where the Saha-Boltzmann factor is defined in \eqref{eq:sabolf}.

\subsection{Emissivities}
The free-bound emissivity
\begin{equation}
 \eta_{ip}^\text{bf} (\nu) = n^*_{i} \alpha_{ip} (\nu)
  \frac{2h\nu^3}{c^2} \exp\zav{-\frac{h\nu}{\bolk \Telec}}.
  \label{ap:bfemis}
\end{equation}
where $\alpha_{ip} (\nu)$ is the photoionization cross section for a transition
from the level $i$ to the level $p$.
 
The free-free emissivity
\begin{equation}
 \label{Eq:etaff}
 \eta^\text{ff}_{i} = n_e N_{j}\alpha^\text{ff}_{i}(\nu, \Telec)
  \frac{2h\nu^3}{c^2}\exp\zav{-\frac{h\nu}{\bolk \Telec}},
\end{equation}
$N_j$ represents the population of the ion $j$, the free-free cross section,
see \citep[Eq. (3.46)]{Kromer2009}
\begin{multline}
 \alpha^\text{ff}_{i_{\text{I}_k}}(\nu, T) = \frac{4e^6}{3hc}
  \left(\frac{2\pi}{3m_\text{e}^2k_\text{B}}\right)^{1/2}
  \frac{(k - 1)^2}{T^{1/2}}\frac{g_\text{ff}(j)}{\nu^3}\\
  = 3.69255\times 10^8 \cdot (k - 1)^2 g_\text{ff}(j)\cdot
   T^{-1/2}\times\nu^{-3}.
\end{multline}
If we put the Eq.~\eqref{Eq:etaff} into Eq.~\eqref{Eq:ff_emission} we can find
easily an analytic formula
\begin{equation}
 \nu = \frac{k_\text{B}T}{h}\log(z),
\end{equation}
where $z\in(0,1)$ is a random number.

\section{Ionisation and excitation equilibrium}
The occupation numbers are in the LTE approximation calculated via the
Boltzmann excitation formula, which we use in the form
\begin{equation}
 \frac{n_{u}}{n_{l}} = \frac{g_{u}}{g_{l}}
  \exp\zav{-\frac{\varepsilon_{u} - \varepsilon_{l}}{\bolk\Telec}},
  \label{Eq:ap:Boltz}
\end{equation}
where $n$ denotes the atomic level number densities, $g$ is the statistical
weight, $\varepsilon$ is the level energy, and the indexes $l$ and $u$ denote
lower and upper level of the transition, respectively. The ionization balance
follows the Saha equation, here in the form
\begin{equation}
 \frac{N_{j}}{N_{p}\nelec} = \frac{U_{j}}{U_{p}}
  \frac{C}{\Telec^{3/2}}
  \exp\zav{\frac{\varepsilon_{p_{\text{L}_0}} - \varepsilon_{j_{\text{L}_0}}}{\bolk\Telec}},
  \label{Eq:ap:Saha}
\end{equation}
where $C$ is the Saha constant
\begin{equation}
 C = \frac{1}{2}\zav{\frac{2\pi\hbar^2}{m_\text{e}\bolk}}^\frac{3}{2},
\end{equation}
$N$ denotes ion number densities, $U$ denotes the partition function,
$\varepsilon$ denotes energies, and the indexes $p$ and $j$ denote atomic
states. The partition function of the ion $i$ is
\begin{equation}
 U_{i}\left(T_\text{e}\right) = \sum\limits^{\mathbf{N}_\text{L}}_{j = 1}
  g_{j} \exp\zav{-\frac{\varepsilon_{j}}{\bolk\Telec}},
  \label{Eq:ap:Part}
\end{equation}
where $\mathbf{N}_\text{L}$ is the total number of included levels for ion $i$.
The sum runs only over the levels included in the calculation.

\end{document}